\documentclass[aps,pra,twocolumn,showpacs,amsmath,amssymb,floatfix]{revtex4-1}

\usepackage{bm}
\usepackage{graphicx}
\usepackage{dcolumn}

\begin{document}

\title{Structural information about the $\text{Ar}_6$ cluster with the
  frozen Gaussian imaginary time propagator}

\author{Holger Cartarius}
\email{Holger.Cartarius@itp1.uni-stuttgart.de}
\affiliation{Institut f\"ur Theoretische Physik 1, Universit\"at Stuttgart,
  70550 Stuttgart, Germany}

\begin{abstract}
  A numerically cheap way to obtain structural information about clusters of
  rare gas atoms at low temperatures is developed. The semiclassical frozen
  Gaussian imaginary time propagator is extended such that it can account for
  the mean values of all inter-atomic distances in the cluster and their
  variances. To reduce the required numerical effort an approximation for
  the mean values is developed which preserves the quality of the results
  offered by the semiclassical ansatz. The method is applied to the
  $\text{Ar}_6$ cluster. It is found that the cluster dissociates almost in
  one step to six free atoms when the temperature is increased. Precursors of
  the dissociation are only observable in the distances of the atoms via the
  appearance of a second isomer. The process is almost classical. However, the
  method is able to resolve small differences in the temperatures at which
  the dissociation takes place and in the mean distances of the bound 
  configuration.
\end{abstract}
\date{\today}

\pacs{32.60.+i, 02.30.-f, 32.80.Fb}

\maketitle

\section{Introduction}

At very low temperatures of a few Kelvin rare gas atoms may assemble and
form clusters due to the van der Waals interaction. It has been found
in a variety of numerical studies \cite{Neirotti00a,Predescu03a,%
  Frantsuzov04a,Predescu05a,White05a,Frantsuzov06a,Perez10a,Adjanor2006a,%
  Pahl2008a,Georgescu2011a,Georgescu2012a} that these clusters exhibit a large
number of interesting effects. In particular, structural changes of the systems
for an increase or decrease of the temperature provide insight into the
behavior and binding mechanisms of the quantum mechanical objects at finite
temperatures. Thus, rare gas atomic clusters are a topic of ongoing research.
Among the rich variety of their thermodynamic properties are a change from one
packing of the atoms to another with increasing energy, or phase transitions
from a solid-like behavior to a liquid-like arrangement of the atoms
\cite{Neirotti00a,Predescu03a,Frantsuzov04a,Predescu05a,White05a,%
  Frantsuzov06a,Perez10a}.

For the weak van der Waals interaction between rare gas atoms the thermal
energy corresponding to a few Kelvin already suffices to lead to all
impacts mentioned above. Thus, accurate quantum mechanical computations are
essential to obtain reliable results. In numerical calculations the Boltzmann
operator at inverse temperature $\beta$, $K = \exp(-\beta H)$, (due to its form
also called imaginary time propagator) is the most important quantity. Its
trace yields the partition function $Z(\beta)$, and the thermal averages of
every observable $O$ follow from  $\langle O \rangle_\beta =
\mathrm{Tr}(K(\beta)O)/Z(\beta)$. However, precise calculations for
multi-dimensional systems are still a challenge for today's numerical
possibilities. For example, path-integral Monte Carlo methods
have been used to investigate rare gas clusters \cite{Berne86a,Makri99a,%
  Ceperley03a}. At low temperatures they become already too expensive for a
few dozen atoms, and efficient but sufficiently accurate approximations are
required. These approximations have to reproduce quantum effects correctly
since they will be of eminent importance at low temperatures. Still many
questions are open. For example $\mathrm{Ne}_{13}$ and $\mathrm{Ne}_{38}$
\cite{Predescu05a,Frantsuzov06a,Adjanor2006a,Pahl2008a} might exhibit novel
low temperature quantum effects, such as liquid-like zero temperature
structures of $\mathrm{Ne}_{38}$ as compared to a solid-like structure
predicted from classical mechanics \cite{Frantz92a}.

Semiclassical methods help even with presently available computation
capabilities to overcome the numerical drawbacks of numerically exact quantum
mechanical algorithms and are still developed and applied in a wide context
concerning thermodynamic properties \cite{Predescu03a,Frantsuzov04a,%
  Predescu05a,Frantsuzov06a,Cartarius11a,Liu2011a,Georgescu2011a,Cartarius12a,%
  Georgescu2012a,Kryvohuz2012a,Conte2012a}. Important semiclassical
approximations are based on the idea to restrict the quantum mechanical wave
functions to a Gaussian shape. The Gaussian functions are defined completely
by a low number of parameters as the position of the center, a momentum of the
wave packet and a width matrix. The lowest numerical effort is achieved with
a frozen Gaussian propagator, i.e.\ a Gaussian wave function of which the
width matrix in the exponent is fixed for all imaginary times (or
temperatures). 

Small argon clusters are among those rare gas clusters which attracted large
interest for a long time and again recently \cite{Etters75a,Leitner89a,%
Leitner91a,Elyutin94a,Gonzales-Lezana99a,Perez10a,Svrckova2011a,Blanco2013a,%
UnnToc2012a,Mella2009a,Calvo2012a,Maehr2007a,Calvo2009a,Franke1993a}.
Despite their apparent simplicity numerical calculations turned out to be
nontrivial, and for the argon trimer even sophisticated path-integral Monte
Carlo calculations could not distinguish a complete dissociation from
structural changes within a bound system because the numerics suffered
strongly from noise \cite{Perez10a}. The frozen Gaussian imaginary time
propagator has proved to provide high quality results. It could solve this
question and identified an almost classical dissociation effect
\cite{Cartarius11a}, which was confirmed \cite{Cartarius12a} with a
first-order correction to the semiclassical imaginary time propagator
\cite{Shao06a,Zhang09a}. 

For the more complicated $\mathrm{Ar}_6$ cluster the specific heat and mean
energy do not provide enough information to understand its whole structure.
It is important to know which alignment of the atoms is present. A very
detailed knowledge of the atomic structure is available via the inter-particle
distances. To obtain their values it is very common to compute the radial pair
correlation function between particles $i$ and $j$ at positions $\bm{r}_i$ and
$\bm{r}_j$, respectively,
\begin{multline}
  p_{ij}(r) = \langle \delta(|\bm{r}_i-\bm{r}_j|-r) \rangle_\beta \\
  = \mathrm{Tr} \left [ K(\beta)  \delta(|\bm{r}_i-\bm{r}_j|-r) \right ]
  /Z(\beta),
  \label{eq:pair_correlation}
\end{multline}
where $\langle \rangle_\beta$ is the thermal average at inverse temperature
$\beta$ and $K(\beta) = \mathrm{e}^{-\beta H}$ represents the imaginary time
propagator at the same temperature \cite{Frantsuzov04a}. This quantity
provides information about the distribution of the distances occurring at
every temperature.

It is the purpose of this paper to show that the frozen Gaussian method
applied to atomic clusters in Ref.\ \cite{Cartarius11a} can be used to
determine the distances of the atoms in clusters in a numerically cheap and
easy way. To do so, an extension to the semiclassical frozen Gaussian method
will be developed. The mean values and the variances of the distances $d_{ij}$
between atoms $i$ and $j$ can directly be accessed and provide a very clear
information of how the atoms are arranged in the cluster. Since the distances
$d_{ij}$ of the single pairs can be calculated in parallel with the same Monte
Carlo sampling all of these values can be obtained with low numerical extra
cost. This provides for $N$ atoms $N(N-1)/2$ independent values, whereas in
the pair correlation function \eqref{eq:pair_correlation} small differences
in the mean distances might be hidden below broad distributions.

The example of the $\mathrm{Ar}_6$ cluster investigated in this article
demonstrates the importance of the additional structural information even in
this relatively simple system, which does not undergo a structural
transformation in the bound state. The mean energy and the specific heat will
show that the cluster dissociates with increasing temperature to a system of
six free atoms in one step. However, precursors of the dissociation will only
be observable in the distances between the atoms, which will indicate a loss
of the ground state configuration at slightly lower temperatures. The
variances of the distances will turn out to be very sensitive to the breakdown
of the structure.

The further sections of this paper are organized as follows. In Sec.\
\ref{sec:method} mean values and variances for inter-atomic distances with
the frozen Gaussian method are introduced. For comparison the same values
are introduced within the more flexible thawed Gaussian variant. Then the 
method is applied to the $\mathrm{Ar}_6$ cluster in Sec.\ \ref{sec:ar6}. After
a short introduction of the system (Sec.\ \ref{sec:system}), the confining
sphere (Sec.\ \ref{sec:sphere}), and the proper choice of the Gaussian
width matrix (Sec.\ \ref{sec:width_matrix}) derivatives of the partition
function (Sec.\ \ref{sec:energy}) and the structural information
(Sec.\ \ref{sec:distances}) are investigated. A discussion in Sec.\
\ref{sec:discussion} concludes the paper.

\section{The frozen Gaussian method}
\label{sec:method}
\allowdisplaybreaks

\subsection{Propagator}
The method is based on a semiclassical approximation of the thermal operator
\begin{equation}
  K(\beta) = e^{-\beta H} ,
  \label{eq:imag_propagator}
\end{equation}
where $\beta = 1/(kT)$ is the inverse temperature. The approximation consists
of evaluating $K(\beta)$ by solving the Bloch equation
\begin{equation}
  -\frac{\partial}{\partial \tau} |\bm{q}_0,\tau \rangle 
  = H  |\bm{q}_0,\tau \rangle
  \label{eq:bloch_equation}
\end{equation}
approximately for a frozen Gaussian coherent state in position space
representation,
\begin{multline}
  \langle \bm{x} | \bm{q}_0(\tau) \rangle 
  = \left ( \frac{\det(\bm{\Gamma})}{\pi^{3N}} \right )^{1/4}
  \exp \biggl ( -\frac{1}{2} [\bm{x} - \bm{q}(\tau)]^\mathrm{T} \\
  \times \bm{\Gamma} 
  [\bm{x} - \bm{q}(\tau)] +\frac{i}{\hbar} \bm{p}^\mathrm{T}(\tau)
  \cdot [\bm{x}-\bm{q}(\tau)]\biggr ) ,
  \label{eq:frozen_state}
\end{multline}
where the width matrix $\bm{\Gamma}$ is a free parameter and has to be adapted
to the given problem as will be explained later for the cluster considered in
this article. The propagation in Eq.\ \eqref{eq:bloch_equation} is done
in imaginary time $\tau$ up to the value $\tau = \beta$ one is interested in.

With Gaussian averages of the type
\begin{equation}
  \langle h(\bm{q}) \rangle =  \int_{-\infty}^{\infty} d\bm{x}^{3N} \,
  |\langle \bm{x} | \bm{q}_0(\tau) \rangle|^2  h(\bm{x})
  \label{eq:Gaverage_general}
\end{equation}
the symmetrized form of the frozen Gaussian propagator is given by
\begin{multline}
  \langle \bm{x}' | K_\mathrm{FG}(\tau) | \bm{x} \rangle 
  = \det(\bm{\Gamma}) \exp \left ( -\frac{\hbar^2}{4} \mathrm{Tr}(\bm{\Gamma})
    \tau \right ) \\ 
  \times \sqrt{\det \left ( 2 \left [ \bm{1} - \exp (-\hbar^2 
        \bm{\Gamma} \tau) \right ]^{-1} \right )} \\
  \times \exp \left ( -\frac{1}{4} [\bm{x}' - \bm{x}]^\mathrm{T} \bm{\Gamma} 
    [\tanh(\hbar^2 \bm{\Gamma} \tau/2)]^{-1} [\bm{x}'-\bm{x}] \right ) \\
  \times \int_{-\infty}^\infty \frac{d\bm{q}^{3N}}{(2\pi)^{3N}} 
  \exp \biggl (-2 \int_0^{\tau/2} d\tau \langle V(\bm{q}(\tau)) \rangle \\
    - [\bm{\bar{x}}-\bm{q}(\tau/2)]^\mathrm{T} \bm{\Gamma} 
    [\bm{\bar{x}}-\bm{q}(\tau/2)] \biggr ) ,
  \label{eq:prop_FG}
\end{multline}
where $\bm{\bar{x}} = (\bm{x}' + \bm{x})/2$. The partition function is simply
given by the trace \cite{Zhang09a}
\begin{multline}
  Z_\mathrm{FG}(\tau) = \mathrm{Tr} \left [ K_\mathrm{FG}(\tau) \right ]
  = \sqrt{\det(\bm{\Gamma})} \exp \left ( -\frac{\hbar^2}{4}
    \mathrm{Tr}(\bm{\Gamma}) \tau \right ) \\
  \times \sqrt{\det \left ( \left [ \bm{1}
        - \exp (-\hbar^2 \bm{\Gamma} \tau) \right ]^{-1} \right )} \\
  \times \int_{-\infty}^\infty \frac{d\bm{q}^{3N}}{(2\pi)^{N/2}} 
  \exp \left (-2 \int_0^{\tau/2} d\tau \langle V(\bm{q}(\tau)) \rangle
  \right ) .
  \label{eq:pf_FG}
\end{multline}
The whole dynamical information is contained in the imaginary time propagation
of the variable $\bm{q}(\tau)$ and is governed by the $3N$ coupled equations
of motion
\begin{equation}
  \frac{\partial \bm{q}(\tau)}{\partial \tau} = -\bm{\Gamma}^{-1}
  \langle \nabla V(\bm{q}(\tau)) \rangle ,
  \label{eq:FG_eqs_motion_q}
\end{equation}
which are relatively simple and can be integrated with a standard integrator
for ordinary differential equations. The remaining numerical task is a single
position space integration for the initial positions $\bm{q}(0)$ in Eqs.\
\eqref{eq:prop_FG} or \eqref{eq:pf_FG}, which is done with a Monte-Carlo
integration.

As shown previously \cite{Cartarius11a,Cartarius12a} a reasonable choice of the
width matrix is crucial for the quality of the semiclassical method. However,
highly precise values can be obtained already with a very simple structure.
Since all particles are identical and thus the pairwise interactions are the
same for all combinations only the center of mass motion has to be
distinguished. In center of mass coordinates,
\begin{subequations}  \begin{align}
    \bm{R}_\mathrm{cm} &= \frac{1}{N} \sum_{i=1}^{N} \bm{r}_i , \\
    \bm{R}_i &= \bm{r}_i - \bm{r}_{i+1} ,  \quad i = 1\,\dots,N-1 ,
  \end{align}
\end{subequations}
good semiclassical estimates are obtained with the matrix
\begin{equation}
  \bm{\Gamma}_\mathrm{cmc} = \begin{pmatrix}
    \bm{D}_1 & \bm{0}   & \cdots \\
    \bm{0}   & \bm{D}_2 & \\
    \vdots   & & \ddots
  \end{pmatrix} 
  \label{eq:gamma_cmc}
\end{equation}
and the $3\times 3$ diagonal matrices $\bm{D}_1$ and $\bm{D}_2$ describing
the three spacial directions of the center of mass and the relative
coordinates, respectively. Since also the spacial directions are equivalent
the best choice are multiples of the unity matrix $\bm{1}_{3\times 3}$, such that
two scalar coefficients $D_1$ and $D_2$ have to be chosen, i.e.
\begin{equation}
  \bm{D}_1 = D_1 \bm{1}_{3\times 3}, \quad \bm{D}_2 = D_2 \bm{1}_{3\times 3} .
  \label{eq:submatrices}
\end{equation}
The form \eqref{eq:prop_FG} of the propagator is represented in Cartesian
coordinates and is the most efficient for the numerical evaluation.
Consequently the Cartesian representation
\begin{subequations}
  \label{eq:FG_general_matrix}
  \begin{equation}
    \bm{\Gamma} = \begin{pmatrix}
      \bar{\bm{D}} & \Delta \bm{D} & \Delta \bm{D} & \\
      \Delta \bm{D} & \bar{\bm{D}} & \Delta \bm{D} &  \cdots \\
      & & \ddots & \\
    \end{pmatrix} ,
  \end{equation}
  \begin{align}
    \bar{\bm{D}} &= (\bm{D}_1+(N-1)\bm{D}_2)/N , \\
    \Delta \bm{D} &= (\bm{D}_1-\bm{D}_2)/N
  \end{align}
\end{subequations}
of the matrix \eqref{eq:gamma_cmc} is used. 

\subsection{Thermal averages of structural information}

Simple thermal averages requiring only the partition function $Z(\beta)$ are
the mean energy $E = \mathrm{k} T^2 \partial \ln Z/\partial T$ and the
specific heat $C = \partial E/\partial T$. However, $K(\beta)$ provides access
to the thermal average of any observable $O$ via
\begin{equation}
  \bar{O}^{\mathrm{(FG)}} = \frac{\mathrm{Tr} (K_\mathrm{FG}(\beta)O)}
  {Z_\mathrm{FG}(\beta)} ,
\end{equation}
which is exploited in this article to gain access to the structural information.
A well suited property is the distance between two atoms, i.e.\
\begin{equation}
  O = d_{ij} = | \bm{x}_i - \bm{x}_j | .
  \label{eq:operator_distances_full}
\end{equation}
With the frozen Gaussian propagator \eqref{eq:prop_FG} this leads to the
expression
\begin{widetext}
  \begin{multline}
    \bar{d}_{ij}^\mathrm{(FG)} = \frac{1}{Z_\mathrm{FG}(\beta)}
    \mathrm{Tr}(K_\mathrm{FG}(\beta) \, | \bm{x}_i - \bm{x}_j |)
    = \frac{1}{Z_\mathrm{FG}(\beta)} \det(\bm{\Gamma}) 
    \exp \left ( -\frac{\hbar^2}{4} \mathrm{Tr}(\bm{\Gamma}) \beta \right )  
    \sqrt{\det \left ( 2 \left [ \bm{1} - \exp (-\hbar^2 \bm{\Gamma} \beta) 
        \right ]^{-1} \right )} \\ \times
    \int_{-\infty}^\infty \frac{\mathrm{d}\bm{q}^{3N}}{(2\pi)^{3N}}  \exp \biggl (-2 
    \int_0^{\beta/2} \mathrm{d}\tau \langle V(\bm{q}(\tau)) \rangle \biggr ) 
    \int_{-\infty}^\infty \mathrm{d}\bm{x}^{3N} \exp \biggl (- [\bm{x}- \bm{q}
    (\beta/2)]^\mathrm{T} \bm{\Gamma} [\bm{x}-\bm{q}(\beta/2)] \biggr )
    | \bm{x}_i - \bm{x}_j | ,
    \label{eq:distance_full_term}
  \end{multline}
\end{widetext}
in which an explicit integration over the $3N$ position variables $\bm{x}$
remains in addition to the evaluation of the partition function
\eqref{eq:pf_FG}. As mentioned previously \cite{Cartarius12a} it is very
important to reduce the numerical effort as much as possible for many-particle
systems. In particular, the position space integrations require an expensive
Monte Carlo sampling in a high-dimensional configuration space.

For usual applications a numerical evaluation of the $\bm{x}$ integration can
be avoided in a reasonable approximation. This can be seen with the variable
$\bm{y} = \bm{x} - \bm{q}(\beta/2)$, which transforms the $\bm{x}$ integral in
\eqref{eq:distance_full_term} to
\begin{multline}
  I = \int_{-\infty}^\infty \mathrm{d}\bm{y}^{3N} \exp \biggl (- \bm{y}^\mathrm{T} 
  \bm{\Gamma} \bm{y} \biggr ) \\ \quad \times
  | \bm{y}_i - \bm{y}_j + \bm{q}_i(\beta/2) - \bm{q}_j(\beta/2)| .
  \label{eq:distance_y}
\end{multline}
The widths of the atom's wave functions contribute only at low temperatures
significantly to the distance. As will be seen, in practical applications a
very narrow Gaussian centers all values $\bm{y}_i$ strongly around zero,
i.e.\ $\bm{x}_i$ is almost identical with $\bm{q}_i(\beta/2)$ for a
nonvanishing Gaussian weight. Thus, the integral \eqref{eq:distance_y} is
calculated for the case $| \bm{y}_i - \bm{y}_j | \ll | \bm{q}_i 
- \bm{q}_j|$. With the expansion
\begin{multline}
  | \bm{y}_i - \bm{y}_j + \bm{q}_i - \bm{q}_j| \approx  | \bm{q}_i - \bm{q}_j |
  - (\bm{y}_i - \bm{y}_j)\cdot \frac{\bm{q}_i - \bm{q}_j}{|\bm{q}_i 
    - \bm{q}_j|} \\
  + \frac{1}{2} (\bm{y}_i - \bm{y}_j)^2 - \frac{1}{2} \frac{\left [ (\bm{y}_i 
      - \bm{y}_j) \cdot (\bm{q}_i - \bm{q}_j ) \right ]^2}{|\bm{q}_i 
    - \bm{q}_j|^3} 
  \label{eq:approximation_dist}
\end{multline}
the integral evaluates to
\begin{multline}
  I = \sqrt{\frac{\pi^{3N}}{\det (\bm{\Gamma})}} \bigg [ |\bm{q}_i(\beta/2) 
  - \bm{q}_j(\beta/2)| \\
  + \frac{\mathrm{Tr} \big ( \bm{\Gamma}_{ii}^{-1} + \bm{\Gamma}_{jj}^{-1}
    - \bm{\Gamma}_{ij}^{-1} - \bm{\Gamma}_{ji}^{-1} \big )}{6 |\bm{q}_i(\beta/2
 ) 
    - \bm{q}_j(\beta/2)|} \bigg ] ,
\end{multline}
where $\bm{\Gamma}_{ij}$ is the $3\times 3$ submatrix of $\bm{\Gamma}$ at the
rows and columns representing particles $i$ and $j$. In total
\begin{widetext}
  \begin{multline}
    \bar{d}_{ij}^\mathrm{(FG)}(\beta) = \frac{1}{Z_\mathrm{FG}(\beta)}
    \mathrm{Tr}(K_\mathrm{FG}(\beta) \, | \bm{x}_i - \bm{x}_j |) 
    \approx \frac{1}{Z_\mathrm{FG}(\beta)} \sqrt{\det(\bm{\Gamma})} 
    \exp \left ( -\frac{\hbar^2}{4} \mathrm{Tr}(\bm{\Gamma}) \beta \right )
    \sqrt{\det \left ( \left [ \bm{1} - \exp (-\hbar^2  \bm{\Gamma} \beta) 
        \right ]^{-1} \right )} \\ \times
    \int_{-\infty}^\infty \frac{\mathrm{d}\bm{q}^{3N}}{(2\pi)^{3N/2}}
    \exp \biggl (-2 \int_0^{\beta/2} \mathrm{d}\tau \langle V(\bm{q}(\tau))
    \rangle \biggr ) \bigg [ |\bm{q}_i(\beta/2) - \bm{q}_j(\beta/2)| 
    + \frac{\mathrm{Tr} \big ( \bm{\Gamma}_{ii}^{-1} + \bm{\Gamma}_{jj}^{-1}
      - \bm{\Gamma}_{ij}^{-1} - \bm{\Gamma}_{ji}^{-1} \big )}{6 |\bm{q}_i(\beta/2)
      - \bm{q}_j(\beta/2)|} \bigg ]
    \label{eq:distance_full_approximated}
  \end{multline}
\end{widetext}
is obtained.

The first term in equation \eqref{eq:distance_full_approximated}, $\propto
|\bm{q}_i - \bm{q}_j|$, reflects the core of the semiclassical approximation,
in which the positions of the atoms are given by the centers $\bm{q}_i$ of the
Gaussian wave packets \eqref{eq:frozen_state}. It corresponds to
\begin{equation}
  O = | \bm{q}_i - \bm{q}_j | .
\end{equation}
The second term contains a correction due to the finite width of an atom's
wave packet. It is completely sufficient to include this lowest-order term, of
which the $\bm{x}$ integration could be done analytically with a simple result,
thus reducing the numerical effort drastically. For the frozen Gaussian method
any higher terms beyond those included in the approximation
\eqref{eq:distance_full_approximated} for the mean distances are of lower
interest. From the physical point of view it is expected that the width of the
atom's wave function only plays a role at very low temperatures at which the
structural configuration is unambiguously in a highly symmetric ground state
configuration. Indeed, as will be seen in the results already the correction
term in the approximation \eqref{eq:distance_full_approximated} is very small. 

To estimate the quality and validity of these mean values additionally the
variances of the distance distributions are calculated. With the operator
\begin{equation}
  O = v_{ij} =  (\bm{x}_i - \bm{x}_j)^2 - {\bar{d}_{ij}}^2 
\end{equation}
and the integral
\begin{widetext}
  \begin{multline}
    \frac{1}{Z_\mathrm{FG}(\beta)}
    \mathrm{Tr} \left ( K_\mathrm{FG}(\beta) \, [ \bm{x}_i - \bm{x}_j ]^2
    \right ) = \frac{1}{Z_\mathrm{FG}(\beta)} \sqrt{\det(\bm{\Gamma})}
    \exp \left ( -\frac{\hbar^2}{4} \mathrm{Tr}(\bm{\Gamma}) \beta \right ) 
    \sqrt{\det \left ( \left [ \bm{1} - \exp (-\hbar^2 \bm{\Gamma} \beta)
        \right ]^{-1} \right )} \\ \times
    \int_{-\infty}^\infty \frac{\mathrm{d}\bm{q}^{3N}}{(2\pi)^{3N/2}}  \exp \biggl
    (-2 \int_0^{\beta/2} \mathrm{d}\tau \langle V(\bm{q}(\tau)) \rangle \biggr ) 
    \bigg [ (\bm{q}_i(\beta/2) - \bm{q}_j(\beta/2))^2 + \frac{1}{2}
    \mathrm{Tr} \big ( \bm{\Gamma}_{ii}^{-1} + \bm{\Gamma}_{jj}^{-1} 
    - \bm{\Gamma}_{ij}^{-1} - \bm{\Gamma}_{ji}^{-1} \big )\bigg ]
    \label{eq:distvariances}
  \end{multline}
\end{widetext}
the standard deviations $\sigma_{ij} = \sqrt{\bar{v}_{ij}^\mathrm{(FG)}}$
of the distances $\bar{d}_{ij}^\mathrm{(FG)}$ are obtained.

\subsection{Sorted distances}
\label{sec:sizeorder}

For the $\mathrm{Ar}_6$ cluster there are 15 possible combinations $i$ and
$j$, and thus 15 distances. The clusters are oriented arbitrarily in
the simulation. The numbers $i$ and $j$ of the atoms have no meaning for
the true configuration, and thus are not appropriate quantities to define the
pairwise distances. The average of all calculations simply results in
identical values for all $\bar{d}_{ij}$, which correspond to the mean value of
all 15 atom-atom distances in a certain configuration. To obtain a meaningful
quantity the distances are sorted according to their size,
\begin{equation}
  d_1 < d_2 < \hdots < d_{15} ,
\end{equation}
and the thermal average of these size-ordered distances is determined, i.e.\
the thermal mean values of the smallest distance, the second smallest, and so
forth are obtained. These values can be compared with the expectations of
geometrical configurations. In an experiment the single distances are
accessible \cite{Kwon1996a} and can in a given sample be sorted the same
way. Alternatively, results from this calculation can be used to determine the
distance of the atoms with a well-grounded assumption about the configuration
\cite{Ulrich2011a}.

\subsection{Comparison with the thawed Gaussian propagator}

The frozen Gaussian method has proved to provide good results for thermodynamic
quantities. We want to know whether or not this is also true for the widths
calculated in this article. Thus, the structural information of the frozen
Gaussian method is compared with that of a more flexible thawed
Gaussian ansatz. It is based on a time-dependent width matrix $\bm{G}(\tau)$,
which adapts itself to the given temperature. With the restriction to Gaussian
wave packets the thawed Gaussian variant is usually the most accurate
approximation. The variable width matrix adds an additional freedom in the
parameters. This is reflected in the quality of the results as has clearly been
demonstrated for a double well potential \cite{Conte10a}. For a large number
of degrees of freedom it suffers, however, from the higher numerical costs. The
single-particle ansatz of Frantsuzov et al. \cite{Frantsuzov04a}
avoids these difficulties by reducing the matrix $\bm{G}(\tau)$ to a
block-diagonal structure, where $3\times 3$ matrices representing the three
spacial coordinates of one particle are the only non-vanishing matrix
elements. In the case of six atoms this reduction is not required and there
is no need to ignore the inter-particle correlations.

The thawed Gaussian propagator used for comparison with the frozen Gaussian
method is the time evolved Gaussian approximation (TEGA) suggested by
Frantsuzov et al. \cite{Frantsuzov03a,Frantsuzov04a} with a full width matrix
$\bm{G}$. It is based on the solution of the Bloch equation
\eqref{eq:bloch_equation} with the coherent state
\begin{multline}
  \langle \bm{x} | g(\bm{q}(\tau),\bm{G}(\tau)) \rangle
  = \left (\pi^{3N} |\det \bm{G}(\tau)|\right )^{-1/4}  \\ \times
  \exp \left ( -\frac{1}{2} [\bm{x}-\bm{q}(\tau)]^\mathrm{T} \bm{G}(\tau)^{-1} 
    [\bm{x}-\bm{q}(\tau)] \right ) .
  \label{eq:tega_gaussian}
\end{multline}
The resulting symmetrized propagator reads
\begin{multline}
  \langle \bm{x} | K_\mathrm{TG}(\tau) | \bm{x}' \rangle   
  = \int \frac{\mathrm{d}\bm{q}^{3N}}{(2\pi)^{3N}} \frac{\exp[2\gamma(\tau/2)]}
  {\det[\bm{G}(\tau/2)]} \\ 
  \times \exp \left ( -\frac{1}{2} [\bm{x}-\bm{q}(\tau/2)
    ]^\mathrm{T} \bm{G}(\tau/2)^{-1} [\bm{x}-\bm{q}(\tau/2)] \right ) \\ \times
  \exp \left ( -\frac{1}{2} [\bm{x}'-\bm{q}(\tau/2)]^\mathrm{T} 
    \bm{G}(\tau/2)^{-1} [\bm{x}'-\bm{q}(\tau/2)] \right )
  \label{eq:prop_TG}
\end{multline}
with the time-dependent width matrix $\bm{G}(\tau)$. In imaginary time $\tau$ 
the equations of motion for the Gaussian parameters $\bm{G}$, $\bm{q}$,
and $\gamma$ are
\begin{subequations}
  \begin{align}
    \frac{d}{d\tau} \bm{G}(\tau) &= -\bm{G}(\tau) \langle \nabla
    \nabla^\mathrm{T} V(\bm{q}(\tau)) \rangle \bm{G}(\tau) + \hbar^2 \bm{1}, 
    \label{eq:tg_eqs_motion_1} \\
    \frac{d}{d\tau} \bm{q}(\tau) &= -\bm{G}(\tau) \langle \nabla V(\bm{q}
    (\tau)) \rangle, \\
    \frac{d}{d\tau} \gamma(\tau) &= -\frac{1}{4} \mathrm{Tr} \left [ \langle
      \nabla\nabla^\mathrm{T} V(\bm{q}(\tau)) \rangle \bm{G}(\tau) \right ] 
    - \langle V(\bm{q}(\tau)) \rangle ,
    \label{eq:tg_eqs_motion_3}
  \end{align}
\end{subequations}
which have to be integrated from $\tau = 0$ to larger times with the initial
conditions
\begin{subequations}
  \begin{align}
    \bm{q}(\tau \approx 0) &= \bm{q}_0 , \\ 
    G(\tau \approx 0) &= \hbar^2 \bm{1} \tau , \label{eq:initial_matrix} \\
    \gamma(\tau \approx 0) &= - V(\bm{q}_0) \tau .
  \end{align}
\end{subequations}
In all expressions $\langle \dots \rangle$ represents Gaussian averaged
quantities of the form \eqref{eq:Gaverage_general} with the wave packet
\eqref{eq:tega_gaussian}, and $\bm{1}$ is the $3N \times 3N$-dimensional
identity matrix. The relevant quantities are the partition function
\begin{equation}
  Z_\mathrm{TG} = \int \frac{\mathrm{d}\bm{q}^{3N}}{(2\sqrt{\pi})^{3N}} 
  \frac{\exp[2\gamma(\tau/2)]}{\sqrt{\det[\bm{G}(\tau/2)]}} 
  \label{eq:pf_TG}
\end{equation}
and the mean value of the distances in the same approximation as for the frozen
Gaussian method,
\begin{multline}
  \bar{d}_{ij}^\mathrm{(TG)}(\beta) \approx \frac{1}{Z_\mathrm{TG}(\beta)}
  \int \frac{\mathrm{d}\bm{q}^{3N}}{(2 \sqrt{\pi})^{3N}}
  \frac{\exp[2\gamma(\beta/2)]}{\sqrt{\det[\bm{G}(\beta/2)]}} \\
  \times \bigg [ |\bm{q}_i(\beta/2) - \bm{q}_j(\beta/2)| \\+ \frac{\mathrm{Tr}
    \big ( \bm{G}_{ii}(\beta/2) + \bm{G}_{jj}(\beta/2) - \bm{G}_{ij}(\beta/2) 
    - \bm{G}_{ji}(\beta/2) \big )}{6 |\bm{q}_i(\beta/2) 
    - \bm{q}_j(\beta/2)|} \bigg ] .
\end{multline}

The thawed Gaussian approximation allows for an additional important
information. Its temperature-dependent width matrix $\bm{G}(\tau)$ provides
easier access to the width of the wave function, which influences the variances
of the distances. The quantum mechanical part of the variances, i.e.\ that
originating from the spread of the wave function, is expected to increase at
lower temperatures. For a frozen Gaussian this can be described
correctly if the constant matrix $\bm{\Gamma}$ is optimized for every single
temperature. In the thawed Gaussian case the variances read
\begin{multline}
  \bar{v}_{ij}^\mathrm{(TG)}(\beta) \approx \frac{1}{Z_\mathrm{TG}(\beta)} 
  \int \frac{\mathrm{d}\bm{q}^{3N}}{(2 \sqrt{\pi})^{3N}}
  \frac{\exp[2\gamma(\beta/2)]}{\sqrt{\det[\bm{G}(\beta/2)]}} \\
  \times \bigg [ (\bm{q}_i(\beta/2) - \bm{q}_j(\beta/2))^2 
  + \frac{1}{2} \mathrm{Tr} \big ( \bm{G}_{ii}(\beta/2) + \bm{G}_{jj}(\beta/2)
  \\ - \bm{G}_{ij}(\beta/2) - \bm{G}_{ji}(\beta/2) \big )\bigg ] 
   - \left ( \bar{d}_{ij}^\mathrm{(TG)} \right )^2 
  \label{eq:var_fctg}
\end{multline}
and follow directly from the imaginary time evolution of $\bm{q}(\tau)$
\emph{and} $\bm{G}(\tau)$. We are mainly interested in the quantum mechanical
part of the variances, viz.\
\begin{multline}
  \bar{v}_{ij}^\mathrm{(TG,qm)}(\beta) \approx \frac{1}{Z_\mathrm{TG}(\beta)} 
  \int \frac{\mathrm{d}\bm{q}^{3N}}{(2 \sqrt{\pi})^{3N}}
  \frac{\exp[2\gamma(\beta/2)]}{\sqrt{\det[\bm{G}(\beta/2)]}} \\
  \times \frac{1}{2} \mathrm{Tr} \big ( \bm{G}_{ii}(\beta/2) 
  + \bm{G}_{jj}(\beta/2) - \bm{G}_{ij}(\beta/2) - \bm{G}_{ji}(\beta/2) \big ) .
   \label{eq:qm_var_fctg}
\end{multline}

\section{Structural information about the
  $\mathrm{Ar}_6$ cluster}
\label{sec:ar6}

\subsection{Representation of the system}
\label{sec:system}

The argon cluster consists of 6 atoms, where the Hamiltonian in mass scaled
coordinates reads
\begin{equation}
  H = -\frac{\hbar^2}{2} \sum_{i=1}^6 \Delta_i 
  + \sum_{j<i} V(r_{ij}) 
  \label{eq:Hamiltonian}
\end{equation}
with the Laplacian $\Delta_i$ of particle $i$. The two-body potential 
$V(r_{ij})$ of Argon is still a very challenging task. One
of the best analytic expressions at hand is a fit to experimental results
by Aziz and Slaman \cite{Aziz86a} of which an adoption to a Morse potential
\cite{Gonzales-Lezana99a} is used,
\begin{equation}
  V(r_{ij}) = D \left ( \exp \left [ -2\alpha (r_{ij}-R_\mathrm{e}) \right ]
    - 2 \exp \left [ -\alpha (r_{ij}-R_\mathrm{e}) \right ]  \right)
  \label{eq:Morse_potential}
\end{equation}
with the parameters $D = 99.00\,\mathrm{cm}^{-1}$, $\alpha = 1.717\,
\text{\r{A}}$, and $R_\mathrm{e} = 3.757\,\text{\r{A}}$ in consistence with
previous studies of the Argon trimer \cite{Perez10a,Cartarius11a,Cartarius12a}.

The numerical efficiency of the frozen Gaussian method is increased with an
expansion of the potential in terms of Gaussians, viz.\
\begin{equation}
  V(|\bm{r}_i - \bm{r}_j|) = \sum_{p} c_p e^{-\alpha_p r_{ij}^2} ,
  \qquad r_{ij} = |\bm{r}_i - \bm{r}_j| .
  \label{eq:Gaussian_fit}
\end{equation}
This procedure was suggested by Frantsuzov et al. \cite{Frantsuzov04a} and
has successfully been applied \cite{Frantsuzov04a,Cartarius11a,Cartarius12a}.
In the form \eqref{eq:Gaussian_fit} Gaussian integrals of the potential or
its derivatives can be done analytically. The required parameters for a fit
to three Gaussians are listed in Table \ref{tab:Gaussian_parameters}
\begin{table}[tb]
  \caption{\label{tab:Gaussian_parameters}Parameters of the Argon-Argon
    interaction potential expressed in terms of Gaussians
    according to Eq.\ \eqref{eq:Gaussian_fit} \cite{Cartarius11a}.}
  \begin{ruledtabular}
    \begin{tabular}{lD{.}{.}{8}D{.}{.}{5}}
      $p$ & \multicolumn{1}{c}{$c_p$ [$\mathrm{cm}^{-1}$]} &
      \multicolumn{1}{c}{$\alpha_p$ [$\text{\r{A}}^{-2}$]} \\
      \hline
      1 & 3.296\times 10^{5} & 0.6551 \\
      2 & -1.279\times 10^{3} & 0.1616 \\
      3 & -9.946\times 10^{3} & 6.0600 \\
    \end{tabular}
  \end{ruledtabular}
\end{table}
and were previously obtained in Ref.\ \cite{Cartarius11a}.

\subsection{Confining sphere}
\label{sec:sphere}

An additional potential is usually introduced to converge the numerical
$\bm{q}$ integration. All particles are confined within a sphere around the
center of mass $\bm{R}_\mathrm{cm}$ by the condition $|\bm{q}-\bm{R}_\mathrm{cm}|
< R_c$, where $R_c$ is the confining radius. This can be achieved with the
steep potential
\begin{equation}
  V_\mathrm{c}(\bm{r}) \propto \sum_{i=1}^{N} \left ( \frac{\bm{r}_i
      - \bm{R}_\mathrm{cm}}{R_\mathrm{c}} \right )^{20} 
  \label{eq:confinement}
\end{equation}
added to the Hamiltonian \eqref{eq:Hamiltonian} or, as in our study, by a
restriction of the volume for the $\bm{q}$ integration.

Of course, an additional potential influences the results and can crucially
change the behavior of the cluster \cite{Predescu03a,Etters75a}. If only bound
configurations are investigated $R_\mathrm{c}$ is usually chosen such that the
bound configurations are not affected, i.e.\ $R_\mathrm{c}$ is larger than the
extension of the bound cluster. However, we are interested also in the
dissociation process for which the choice of $R_\mathrm{c}$ is nontrivial 
\cite{Cartarius11a,Cartarius12a}. A larger radius $R_\mathrm{c}$ always allows
for a dissociation at lower temperatures. In principle it has to be adopted
to the physical conditions as, e.g.\ the pressure. We are interested in
the qualitative behavior at the dissociation and it was checked carefully that
the choice $R_\mathrm{c} = 35\,\text{\r{A}}$ does not influence the qualitative
change of the relevant observables, i.e.\ the mean energy, the specific heat,
the mean values of the inter-atomic distances and their variances. In
particular, it was assured that the case of a completely dissociated cluster
is present for temperatures above $40\,\mathrm{K}$ and the form of the
dissociation process is not altered. The value of the confining radius
$R_\mathrm{c}$, i.e.\ the pressure in physical terms, affects the temperature
at which the dissociation occurs.

\subsection{Choice of the width matrix}
\label{sec:width_matrix}

While in a thawed Gaussian calculation the initial condition for the width
matrix \eqref{eq:initial_matrix} is defined, the constant matrix $\bm{\Gamma}$
of its frozen Gaussian counterpart has to be chosen carefully. It is a free
parameter of the system. It is not trivial to find a good choice of
$\bm{\Gamma}$. However, as was mentioned above, the structure
\eqref{eq:FG_general_matrix} with the $3\times 3$ submatrices
\eqref{eq:submatrices} is well suited. Thus, only the two parameters $D_1$ and
$D_2$ need to be chosen.

A detailed investigation of the $\mathrm{Ar}_3$ cluster revealed that there is
a reliable and simple method to find the best choice for the inter-particle
width parameter $D_2$ \cite{Cartarius11a}. Propagating the partition function
\eqref{eq:pf_FG} to long imaginary times $\beta \to \infty$, i.e.\ $T \to 0$,
one can extract the thermodynamic mean energy to correspond to the ground
state energy $E_0$. The parameter $D_2$ providing the lowest value for $E_0$
has shown to lead to the best agreement with numerically exact calculations
and the more flexible thawed Gaussian approximation. This result is almost
independent of the temperature at which the partition function, mean energy or
specific heat of all methods are compared. Furthermore, calculating the
first-order correction to the frozen Gaussian propagator \eqref{eq:prop_FG}
showed that this choice also requires the smallest correction. Thus, the
simple minimization of the ground state energy gives us a reliable way of
determining $D_2$. For the $\mathrm{Ar}_6$ cluster in this article it was found
that $D_2 = 32\,\text{\r{A}}^{-2}$ is the best choice.

The center of mass is free and it can exactly be described by a Gaussian
in the limit $D_1 \to 0$. This means the value should be as small as possible.
For the numerical evaluation one needs a finite value. It is known 
that $D_1 = 0.1\,\text{\r{A}}^{-2}$ is small enough \cite{Cartarius11a}. The
results cannot be distinguished from those of even lower values for $D_1$.

\subsection{Mean energy and specific heat}
\label{sec:energy}

\begin{figure}[tbp]
  \centering
  \includegraphics[width=\columnwidth]{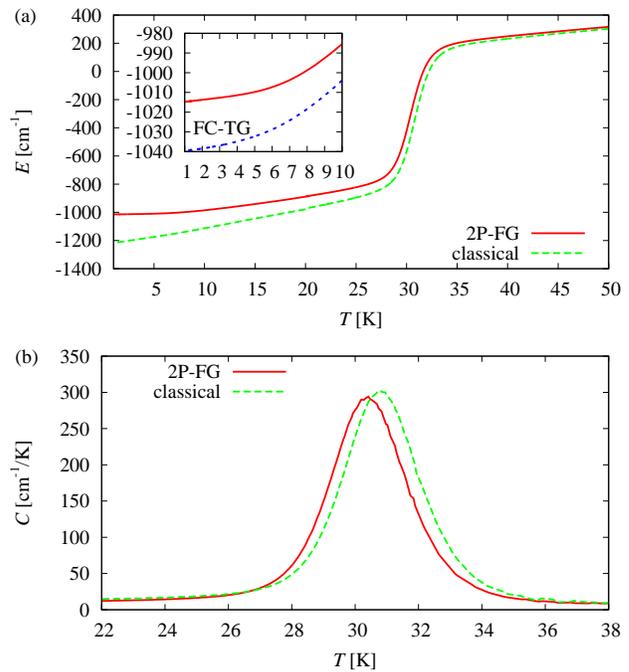}
  \caption{\label{fig:energy}(Color online) (a) Mean energies of the
    $\mathrm{Ar}_6$ cluster calculated with the two-parameter frozen Gaussian
    propagator (2P-FG) and its classical counterpart. For the high-temperature
    limit both results agree well. At low temperatures the classical mean
    energy tends to the potential minimum and the frozen Gaussian propagator
    approximates the quantum mechanical ground state. The inset shows a
    comparison of the 2P-FG method with the fully coupled thawed Gaussian
    method (FC-TG). The energies differ by a few percent. (b) Specific heat
    around the dissociation, which seems to happen in one step.}
\end{figure}

To compare the $\mathrm{Ar}_6$ cluster with the trimer considered in
\cite{Cartarius11a,Cartarius12a} the mean energy and the specific heat
are studied first. They are shown in Fig.\ \ref{fig:energy}
for the two-parameter ($D_1$ and $D_2$) frozen Gaussian propagator and the
derivatives of the classical partition function
\begin{equation}
  Z_\mathrm{cl} = \left ( \frac{\mathrm{k} T}{2\pi \hbar^2} \right )^{3/2 N}
  \int \mathrm{e}^{-\beta V(\bm{q})} \, \mathrm{d}\bm{q}^{3N} .
  \label{eq:pf_classical}
\end{equation}
The observations are very similar to those obtained for trimer. At very low
temperatures the classical calculation tends to the potential minimum. At
$T = 1\,\mathrm{K}$ a mean energy of $E = -1216\,\mathrm{cm}^{-1}$ is found.
The frozen Gaussian results are at this temperature already in a very flat
regime, in which the mean energy is almost independent of the temperature and
approximates the quantum mechanical ground state energy. The method leads to
$E_0 \approx -1015\, \mathrm{cm}^{-1}$. In the inset of Fig.\ \ref{fig:energy}
(a) a comparison with the more flexible fully-coupled thawed Gaussian
propagator is shown. It leads to a value of $E_0 \approx -1040\,
\mathrm{cm}^{-1}$, i.e.\ the difference of the ground state's binding energy
is only $2.4\%$. Thus, one may conclude that also for the larger
$\mathrm{Ar}_6$ cluster the quality of the frozen Gaussian propagator is
acceptable in comparison with the numerically more expensive thawed Gaussian
variant even for the low-temperature limit.

The dissociation appears in the mean energy as a step. The energy raises
almost directly to that of six free particles. This indicates a dissociation
of all atoms at once as was observed for the trimer. The same information
can be gained from the specific heat, which is shown around the dissociation
in Fig.\ \ref{fig:energy} (b). One broad peak confirms that the dissociation
occurs in one step. The classical calculation shows a transition at a slightly
lower temperature, and the difference between the two maxima in the specific
heat is approximately $0.5\,\mathrm{K}$, which is lower than for the trimer,
where a difference of $1.5\,\mathrm{K}$ was observed. Certainly the difference
can again be related to the zero point energy in quantum mechanics. This is
larger for six atoms than for three and one could expect that also the
temperature difference is larger. However, one has to keep in mind that this
energy has to be distributed among a larger number of atoms during the
dissociation. Since aside from the small shift in the temperature the
dissociation process is almost identical in the classical and the frozen
Gaussian calculation one may conclude that it is a purely classical phenomenon.

\subsection{Structural information for low temperatures and
  for the dissociation}
\label{sec:distances}

In addition to the information of the simple derivatives of the partition
function the mean distances are studied. Since so far the dissociation
seems to be purely classical it is interesting to also compare the structural
information with the classical one. To do so, the classical mean distances 
\begin{equation}
  \bar{d}_{ij}^\mathrm{(classical)} = \frac{1}{Z_\mathrm{cl}} 
  \left ( \frac{\mathrm{k} T}{2\pi \hbar^2} \right )^{3/2 N}
  \int \mathrm{e}^{-\beta V(\bm{q})} | \bm{q}_i - \bm{q}_j| \, 
  \mathrm{d}\bm{q}^{3N}
  \label{eq:mean_dist_classical}
\end{equation}
are added in the calculations below.

\subsubsection{Structure at low temperatures}

\begin{figure}[tbp]
  \centering
  \includegraphics[width=\columnwidth]{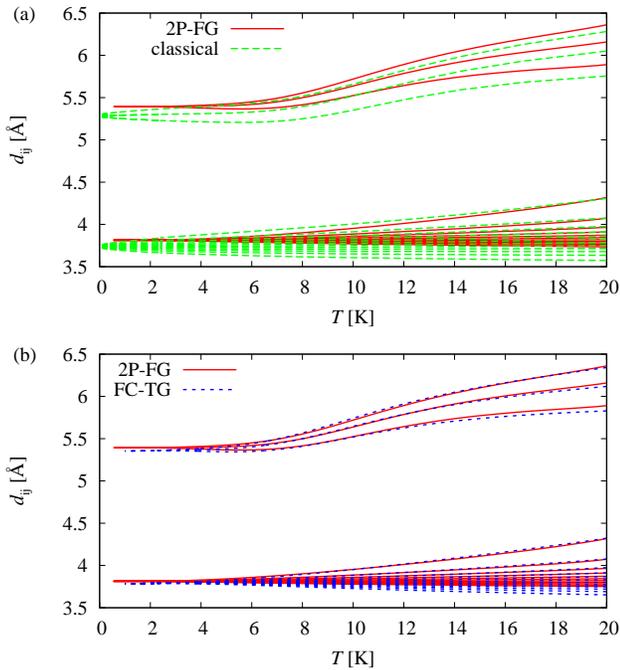}
  \caption{\label{fig:dist_low}(Color online) (a) Comparison of the mean values
    of all 15 distances calculated with the frozen Gaussian method (2P-FG)
    and a classical calculation at temperatures $T \leq 20 $. The distances
    appear in groups. Three distances converge for $T\to 0$ to values above
    $5\,\text{\r{A}}$, whereas the remaining 12 are below $4\,\text{\r{A}}$.
    (b) A comparison of the frozen Gaussian and thawed Gaussian method (FC-TG)
    shows that the mean distances agree very well.}
\end{figure}

The mean distances obtained for temperatures below $20\,\mathrm{K}$, i.e.\
significantly below the dissociation process, are shown in Fig.\
\ref{fig:dist_low},
where first the frozen Gaussian method (2P-FG) is compared with the classical
calculation in Fig.\ \ref{fig:dist_low} (a), and then the fully-coupled
thawed Gaussian approximation is added in Fig.\ \ref{fig:dist_low}
(b). The most striking observation is that the distances appear in the
low-temperature limit in two groups. A group of three ``long'' distances,
of which the values are always above $5\,\text{\r{A}}$ for $T\to 0$, and
a second group of the 12 remaining ``short'' distances, which converge to a
value below $4\,\text{\r{A}}$, exist. This already gives a clear answer to the
question about the ground state configuration of $\mathrm{Ar}_6$. It is
consistent with the distances in an octahedron, or in other words, the atoms
are located at the centers of the surfaces of its dual polyhedron, viz.\
the cube. Twelve short distances $d_s$ from the atoms on neighboring surfaces
and three longer distances $d_l$ between the atoms on opposite surfaces are
expected. The ratio of the distances is supposed to be $d_l = \sqrt{2} d_s$,
which is fulfilled excellently in both the classical and semiclassical
calculations.

In the classical case the atoms seek directly the potential minima, whereas
in the quantum case always a wave function with a finite width leading
automatically to larger mean distances is present. Furthermore, in the classical
case the fixed octahedron configuration is only observable for $T\to 0$.
This is a consequence of the fact that classically every nonvanishing energy
allows for a thermal excitation. In contrast to this there should be no
excitation possible if $\mathrm{k}T$ is clearly below the energy difference
between the ground state and the first excited state in the quantum mechanical
case. This is also reflected in the mean distances. For temperatures below
$T\approx 3\,\mathrm{K}$ no differences between the distances in one group
are observed, and the distances do not change for even lower temperatures.
This indicates that the cluster is already in the ground state configuration.

A comparison of the frozen Gaussian method with the fully-coupled thawed
Gaussian propagator reveals that the distances agree very well. This is
in particular true for all larger distances. Also the low temperature
limit shows an excellent agreement. The edge length of the cube containing
the octahedron is $d_l = 5.39\,\text{\r{A}}$ in the frozen Gaussian
calculation and $d_l = 5.35\,\text{\r{A}}$ in the thawed Gaussian
approximation. The difference is below $1\,\%$, and thus even smaller than that
of the mean energy. Obviously the structural information of the frozen Gaussian
method is less affected by the constant Gaussian width approximation.

With the data of Fig.\ \ref{fig:dist_low} we are also able to estimate
the quality of the approximation \eqref{eq:approximation_dist}, in which
the power series expansion of the distances was introduced. The first-order
term retained in the expansion is of the size
\begin{equation*}
  \frac{\mathrm{Tr} \big ( \bm{\Gamma}_{ii}^{-1} + \bm{\Gamma}_{jj}^{-1}
    - \bm{\Gamma}_{ij}^{-1} - \bm{\Gamma}_{ji}^{-1} \big )}{6 |\bm{q}_i 
    - \bm{q}_j|} ,
\end{equation*} 
where for the width matrix \eqref{eq:FG_general_matrix} 
$\mathrm{Tr} \big ( \bm{\Gamma}_{ii}^{-1} + \bm{\Gamma}_{jj}^{-1} 
- \bm{\Gamma}_{ij}^{-1} - \bm{\Gamma}_{ji}^{-1} \big )/6 = D_2^{-1} 
= (32\,\text{\r{A}}^{-2})^{-1} = 0.031\,\text{\r{A}}^{2}$ is obtained.
With the knowledge that the typical distances $|\bm{q}_i - \bm{q}_j|$ are
even in the bound phase of the order of a few \r{A}ngstr\"oms this correction
can be estimated to be always less than $1\,\%$ of the leading order. Hence,
it has at most the same size as the difference between the two semiclassical
propagators. Higher orders in the series expansion
\eqref{eq:approximation_dist} would lead to even smaller corrections, which do
not need to be taken into account since they are below the error of the
semiclassical approximation.

\subsubsection{Dissociation to six free atoms}

\begin{figure}[tbp]
  \centering
  \includegraphics[width=\columnwidth]{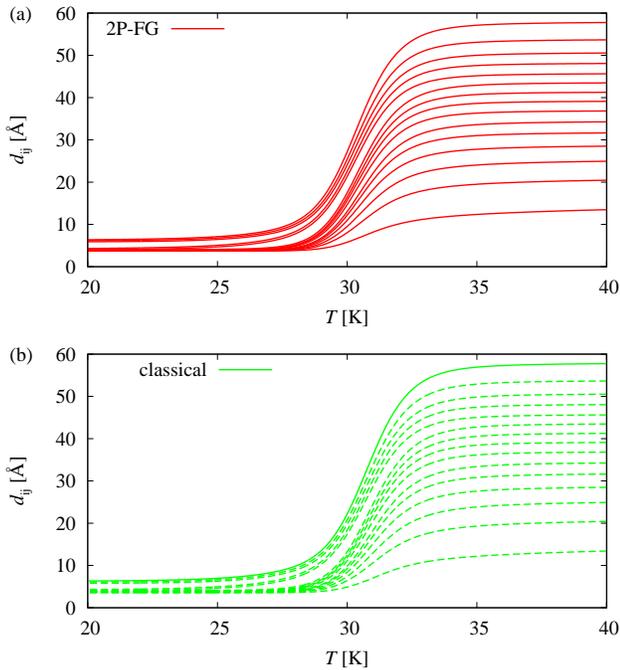}
  \caption{\label{fig:dist_high}(Color online) (a) Mean values of all 15
    distances calculated with the frozen Gaussian method (2P-FG) around the
    dissociation. Two of the smaller distances join the group of the three
    larger distances starting at $T \approx 23\,\mathrm{K}$, but then the
    dissociation occurs at once. (b) The same behavior is observed for the
    classical calculation.}
\end{figure}

The distances around the dissociation are shown in Fig.\ \ref{fig:dist_high},
where the classical and 2P-FG results are compared. Since the frozen Gaussian
approximation is known to provide good results at these temperatures (cf.\
Ref.\ \cite{Cartarius12a}), a comparison with the thawed Gaussian propagator
does not give any new information. Figure \ref{fig:dist_high} confirms the
finding of the consideration of the mean energy and the specific heat in
Fig.\ \ref{fig:energy}. The dissociation effect is classical. The
semiclassical approximation of the quantum mechanical propagator and the
purely classical calculation lead to the same behavior. Apart from a small
shift in temperature both diagrams agree very well.

\begin{figure}[tbp]
  \centering
  \includegraphics[width=\columnwidth]{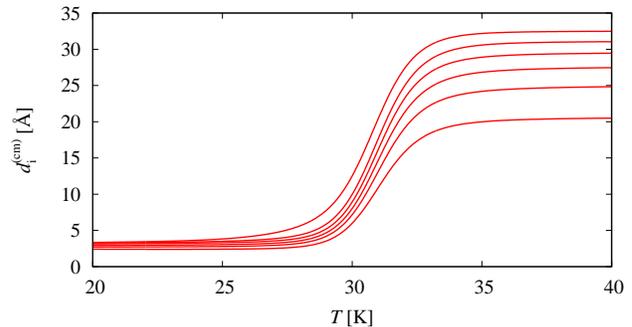}
  \caption{\label{fig:cmdist}(Color online) Distances of all six atoms from
    the center of mass in a classical calculation for $R_\mathrm{c} = 35\,
    \text{\r{A}}$. Shown is a thermal average of size-ordered distances as
    explained in Sec.\ \ref{sec:sizeorder}. With increasing temperature one
    distance becomes larger than the others before all raise drastically.}
\end{figure}

The calculation of the distances gives additional insight into the dissociation
process. For temperatures $T \lessapprox 23\,\mathrm{K}$ the groups of short
and long distances are unchanged. Above this temperature two of the twelve
distances $d_s$ are separating from the others and join the three longer
distances $d_l$. Two new groups with five and ten distances start to from.
For $\text{Ar}_6$ a second isomer in the form of a tri-tetrahedron is known
to contribute at increasing temperatures \cite{Franke1993a}. This would
exactly agree with a grouping of five longer and ten shorter distances and
is also visible in Fig.\ \ref{fig:cmdist}, in which the distances of all six
atoms from the center of mass are shown for a classical calculation.
At low temperatures all distances have almost the same size, which agrees with
a pure octahedron configuration. With increasing temperature one distance
grows and indicates a coexistence of octahedron and tri-tetrahedron
configurations. Signatures of further arrangements of the atoms are not
found.

However, this rearrangement of the atoms remains in its beginnings. A new
structure cannot completely arise since the whole process does not finish
before all distances raise drastically and indicate with this increase the
dissociation of the cluster. Comparisons with calculations, in which due
to a smaller value of $R_c$ no dissociation is allowed, show that this effect
only appears in connection with the dissociation. Thus, the separation of the
two distances is more a precursor of the total destruction of the cluster.
The dissociation happens then at once. After the dissociation the distances
obtain new almost constant values which correspond to the distribution of
atoms moving freely within the confining sphere.

\subsubsection{Variances of the distances}

\begin{figure}[tbp]
  \centering
  \includegraphics[width=\columnwidth]{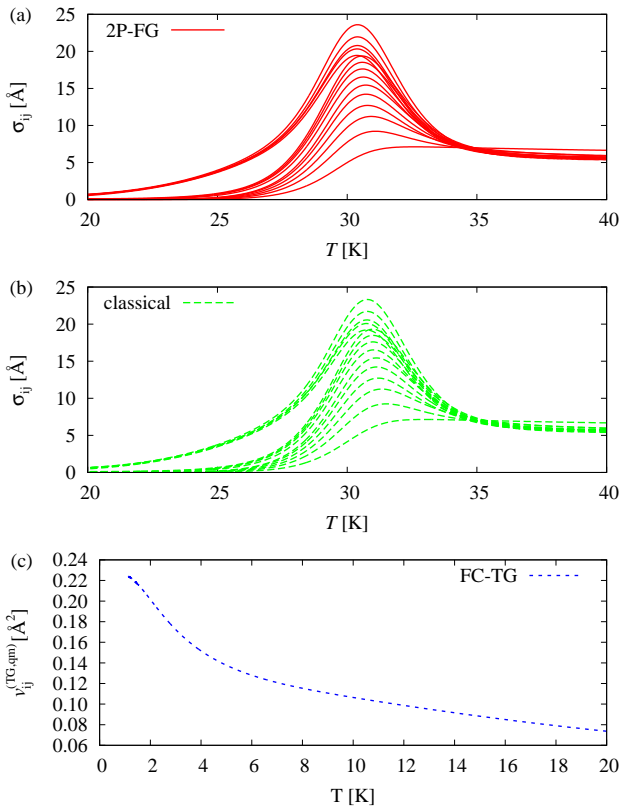}
  \caption{\label{fig:variances}(Color online) (a) Standard deviations
    $\sigma_{ij}$ of the distances in the frozen Gaussian approximation (2P-FG)
    and (b) of the classical calculation. They increase drastically around the
    dissociation. (c) Quantum mechanical part \eqref{eq:qm_var_fctg} of
    the variances for the fully coupled thawed Gaussian propagator. For low
    temperatures it increases due to an increasing width of the Gaussians.}
\end{figure}

To learn more about the actual distribution of the distances their standard
deviations are shown in Figs.\ \ref{fig:variances} (a) and (b)
for the classical and the 2P-FG method. The most significant feature is the
drastic increase of the standard deviations around the transition. At this
temperature range parts of the simulated clusters are still in a bound
configuration whereas others are already dissociated. The effect is similar
and of the same size for the classical and the semiclassical calculation.
For temperatures above the dissociation the standard deviations are almost
the same for all distances, which is also expected for six free atoms.

Of more interest is the behavior of the standard deviations below the
dissociation. The longer distances are expected to show more fluctuations.
Additionally two of the shorter distances join the group of the three longer
distances for increasing temperatures as a consequence of contributions from
two isomers. It can also be expected that these two show higher standard
deviations than the short distances since the separation of the two distances
does not happen abruptly at one temperature as can be seen in Fig.\
\ref{fig:dist_high}. Consequently, below the dissociation the standard
deviations are found to form two groups. One group combines the standard
deviations of five distances, i.e.\ the three longer ones and the two joining
them. The other group consists of the standard deviations of the 10 short
distances which stay together up to the temperature of the dissociation.

In Figs.\ \ref{fig:variances} (a) and (b) it seems that the standard deviations
always decrease for lower temperatures. This is definitely expected for
the classical calculation. The frozen Gaussian approximation can, since the
width of the wave function is determined by the constant values of
$\bm{\Gamma}$, not reflect the quantum mechanical expectation that the wave
function stretches at lower temperatures. To cover also this effect the
quantum mechanical part according to equation \eqref{eq:qm_var_fctg} is
plotted in Fig.\ \ref{fig:variances} (c). Only the variance of one of the
distances is shown since this part is almost identical for all of them in the
highly symmetric situation of a monoatomic cluster. At $T \approx
20\,\mathrm{K}$ the variance has approximately the same size as that following
from the frozen Gaussian method $v_{ij}^\mathrm{(FG,qm)} = \mathrm{Tr}
\big ( \bm{\Gamma}_{ii}^{-1} + \bm{\Gamma}_{jj}^{-1} - \bm{\Gamma}_{ij}^{-1} 
- \bm{\Gamma}_{ji}^{-1} \big )/2 = 3 D_2^{-1} = 0.093\,\text{\r{A}}^2$. In
particular, these quantum mechanical parts of the variances are considerably
lower than other contributions in equations \eqref{eq:distvariances} and
\eqref{eq:var_fctg}. Figure \ref{fig:variances} (a) would not change with the
thawed Gaussian propagator. Only for temperatures $T \lessapprox 10\,
\mathrm{K}$ the extension of the wave function becomes important for a
measurement of the distances. For temperatures in the range of the
dissociation the information from the frozen Gaussian method is completely
sufficient.

All calculations shown in this section could be implemented and performed
easily on a NVIDIA Tesla C2070 GPU. On this architecture a converged result
for low temperatures is obtained in less than 20 hours. The most critical part
is that around the dissociation, where a very detailed sampling for bound
configurations has to be done alongside an inclusion of large distances
allowing for an unbound cluster. This can require an increase of the sampling
points by a factor of 10.

\section{Summary and outlook}
\label{sec:discussion}

In this article it was shown that structural information about a cluster of
atoms can be obtained with the frozen Gaussian semiclassical method in a
numerically cheap way. The evaluation of the corresponding integrals can be
done in parallel to that of the partition function. With this method the full
information about all the distances of all combinations of the atoms can be
obtained. A comparison with the more flexible thawed Gaussian propagator
revealed that the quality of the distances is on the same level as that of the
mean energy or the specific heat, or even better. To avoid inefficient
numerical computations of a position space integral an approximation for the
distances was introduced. It was possible, however, to show that this
approximation does not reduce the quality of the results below that obtained
in the semiclassical approximation of the propagator.

On the physical side it was found that with increasing temperatures the
$\mathrm{Ar}_6$ cluster undergoes an almost direct transition to six free
atoms. However, it shows precursors in the distances. At temperatures
slightly below the dissociation a reordering of the atoms starts, in which
contributions from a second isomer, viz.\ a tri-tetrahedron \cite{Franke1993a}, 
appear, but then vanish in the increasing distances at the dissociation. The
dissociation is a purely classical effect. The semiclassical approximation
shows exactly the same behavior with just a small shift in the temperature of
$0.5\,\mathrm{K}$. Around the dissociation the standard deviations of the
distances are almost completely determined by classical contributions. Only
for lower temperatures the extension of the wave functions becomes important
as was seen in a thawed Gaussian approximation.

At low temperatures the cluster assumes the shape of an octahedron, where the
longer distance between the atoms is $d_l = 5.4\,\text{\r{A}}$, and the shorter
has the value $d_s = d_l/\sqrt{2} = 3.8\,\text{\r{A}}$. Classically the fixed
configuration is only obtained in the limit $T \to 0$ whereas in the quantum
mechanical case the ground state configuration is present for all temperatures 
$T\lessapprox 3\,\mathrm{K}$. 

The frozen Gaussian method has proved to provide reliable results for quantum
mechanical calculations. There is a large number of investigations which can
be done with it. In particular, the results for $\mathrm{Ar}_3$ in
\cite{Cartarius11a} and for $\mathrm{Ar}_6$ in this work indicate that the
confinement to very small spheres usually applied in the calculation of the
partition function and values deduced from it
\cite{Predescu03a,Frantsuzov04a,Predescu05a,Frantsuzov08a} only sample
bound cluster configurations. This is physically realized at high pressures.
If one is interested in lower pressures, at which a dissociation is
allowed, this is too restrictive to fully understand the low-temperature
behavior of the clusters. The dissociation can set in before structural changes
or a melting can be observed. To take this into account it is necessary to
advance the investigations done here to clusters with higher numbers of atoms.
In particular, the cases of $\mathrm{Ar}_{13}$ \cite{Franke88a,Borrmann94a,%
Tsai93a}, $\mathrm{Ne}_{13}$ \cite{Frantsuzov04a} or $\mathrm{Ne}_{38}$
\cite{Predescu05a} are of special interest since they showed interesting
structural transformations in the non-dissociated cases. Whereas the stronger
quantum effects in the completely bound case are well covered by a variable
width matrix the numerically cheaper frozen Gaussian method has advantages
in the numerically more challenging case of the dissociation requiring
a sampling of bound and unbound configurations of the atoms. The exactness
of both methods can then be monitored and (if necessary) improved with the the
series expansion of the imaginary time propagator
\cite{Shao06a,Zhang09a,Cartarius12a}.

The importance of the series expansion is not restricted to the dissociation.
Most effects in rare gas clusters such as structural transformations or
dissociations appear at such low temperatures that it is necessary to analyze
whether the semiclassical approximations used in the calculations correctly
reproduce the true quantum mechanical behavior. An important example will be
$\mathrm{Ne}_{38}$, for which strong differences are found between the
approximate quantum computations and a purely classical theory
\cite{Frantsuzov06a}.

\begin{acknowledgments}
  H.C. is grateful for a Minerva fellowship. He thanks Eli Pollak for valuable
  comments and kind hospitality at the Weizmann Institute of Science, where
  this work has been started.
\end{acknowledgments}


\begin{thebibliography}{44}%
\makeatletter
\providecommand \@ifxundefined [1]{%
 \@ifx{#1\undefined}
}%
\providecommand \@ifnum [1]{%
 \ifnum #1\expandafter \@firstoftwo
 \else \expandafter \@secondoftwo
 \fi
}%
\providecommand \@ifx [1]{%
 \ifx #1\expandafter \@firstoftwo
 \else \expandafter \@secondoftwo
 \fi
}%
\providecommand \natexlab [1]{#1}%
\providecommand \enquote  [1]{``#1''}%
\providecommand \bibnamefont  [1]{#1}%
\providecommand \bibfnamefont [1]{#1}%
\providecommand \citenamefont [1]{#1}%
\providecommand \href@noop [0]{\@secondoftwo}%
\providecommand \href [0]{\begingroup \@sanitize@url \@href}%
\providecommand \@href[1]{\@@startlink{#1}\@@href}%
\providecommand \@@href[1]{\endgroup#1\@@endlink}%
\providecommand \@sanitize@url [0]{\catcode `\\12\catcode `\$12\catcode
  `\&12\catcode `\#12\catcode `\^12\catcode `\_12\catcode `\%12\relax}%
\providecommand \@@startlink[1]{}%
\providecommand \@@endlink[0]{}%
\providecommand \url  [0]{\begingroup\@sanitize@url \@url }%
\providecommand \@url [1]{\endgroup\@href {#1}{\urlprefix }}%
\providecommand \urlprefix  [0]{URL }%
\providecommand \Eprint [0]{\href }%
\providecommand \doibase [0]{http://dx.doi.org/}%
\providecommand \selectlanguage [0]{\@gobble}%
\providecommand \bibinfo  [0]{\@secondoftwo}%
\providecommand \bibfield  [0]{\@secondoftwo}%
\providecommand \translation [1]{[#1]}%
\providecommand \BibitemOpen [0]{}%
\providecommand \bibitemStop [0]{}%
\providecommand \bibitemNoStop [0]{.\EOS\space}%
\providecommand \EOS [0]{\spacefactor3000\relax}%
\providecommand \BibitemShut  [1]{\csname bibitem#1\endcsname}%
\let\auto@bib@innerbib\@empty
\bibitem [{\citenamefont {Neirotti}\ \emph {et~al.}(2000)\citenamefont
  {Neirotti}, \citenamefont {Freeman},\ and\ \citenamefont
  {Doll}}]{Neirotti00a}%
  \BibitemOpen
  \bibfield  {author} {\bibinfo {author} {\bibfnamefont {J.~P.}\ \bibnamefont
  {Neirotti}}, \bibinfo {author} {\bibfnamefont {D.~L.}\ \bibnamefont
  {Freeman}}, \ and\ \bibinfo {author} {\bibfnamefont {J.~D.}\ \bibnamefont
  {Doll}},\ }\href@noop {} {\bibfield  {journal} {\bibinfo  {journal} {J. Chem.
  Phys.}\ }\textbf {\bibinfo {volume} {112}},\ \bibinfo {pages} {3990}
  (\bibinfo {year} {2000})}\BibitemShut {NoStop}%
\bibitem [{\citenamefont {Predescu}\ \emph {et~al.}(2003)\citenamefont
  {Predescu}, \citenamefont {Sabo}, \citenamefont {Doll},\ and\ \citenamefont
  {Freeman}}]{Predescu03a}%
  \BibitemOpen
  \bibfield  {author} {\bibinfo {author} {\bibfnamefont {C.}~\bibnamefont
  {Predescu}}, \bibinfo {author} {\bibfnamefont {D.}~\bibnamefont {Sabo}},
  \bibinfo {author} {\bibfnamefont {J.~D.}\ \bibnamefont {Doll}}, \ and\
  \bibinfo {author} {\bibfnamefont {D.~L.}\ \bibnamefont {Freeman}},\
  }\href@noop {} {\bibfield  {journal} {\bibinfo  {journal} {J. Chem. Phys.}\
  }\textbf {\bibinfo {volume} {119}},\ \bibinfo {pages} {12119} (\bibinfo
  {year} {2003})}\BibitemShut {NoStop}%
\bibitem [{\citenamefont {Frantsuzov}\ and\ \citenamefont
  {Mandelshtam}(2004)}]{Frantsuzov04a}%
  \BibitemOpen
  \bibfield  {author} {\bibinfo {author} {\bibfnamefont {P.~A.}\ \bibnamefont
  {Frantsuzov}}\ and\ \bibinfo {author} {\bibfnamefont {V.~A.}\ \bibnamefont
  {Mandelshtam}},\ }\href@noop {} {\bibfield  {journal} {\bibinfo  {journal}
  {J. Chem. Phys.}\ }\textbf {\bibinfo {volume} {121}},\ \bibinfo {pages}
  {9247} (\bibinfo {year} {2004})}\BibitemShut {NoStop}%
\bibitem [{\citenamefont {Predescu}\ \emph {et~al.}(2005)\citenamefont
  {Predescu}, \citenamefont {Frantsuzov},\ and\ \citenamefont
  {Mandelshtam}}]{Predescu05a}%
  \BibitemOpen
  \bibfield  {author} {\bibinfo {author} {\bibfnamefont {C.}~\bibnamefont
  {Predescu}}, \bibinfo {author} {\bibfnamefont {P.~A.}\ \bibnamefont
  {Frantsuzov}}, \ and\ \bibinfo {author} {\bibfnamefont {V.~A.}\ \bibnamefont
  {Mandelshtam}},\ }\href@noop {} {\bibfield  {journal} {\bibinfo  {journal}
  {J. Chem. Phys.}\ }\textbf {\bibinfo {volume} {122}},\ \bibinfo {eid}
  {154305} (\bibinfo {year} {2005})}\BibitemShut {NoStop}%
\bibitem [{\citenamefont {White}\ \emph {et~al.}(2005)\citenamefont {White},
  \citenamefont {Cleary},\ and\ \citenamefont {Mayne}}]{White05a}%
  \BibitemOpen
  \bibfield  {author} {\bibinfo {author} {\bibfnamefont {R.~P.}\ \bibnamefont
  {White}}, \bibinfo {author} {\bibfnamefont {S.~M.}\ \bibnamefont {Cleary}}, \
  and\ \bibinfo {author} {\bibfnamefont {H.~R.}\ \bibnamefont {Mayne}},\
  }\href@noop {} {\bibfield  {journal} {\bibinfo  {journal} {J. Chem. Phys.}\
  }\textbf {\bibinfo {volume} {123}},\ \bibinfo {eid} {094505} (\bibinfo {year}
  {2005})}\BibitemShut {NoStop}%
\bibitem [{\citenamefont {Frantsuzov}\ \emph {et~al.}(2006)\citenamefont
  {Frantsuzov}, \citenamefont {Meluzzi},\ and\ \citenamefont
  {Mandelshtam}}]{Frantsuzov06a}%
  \BibitemOpen
  \bibfield  {author} {\bibinfo {author} {\bibfnamefont {P.~A.}\ \bibnamefont
  {Frantsuzov}}, \bibinfo {author} {\bibfnamefont {D.}~\bibnamefont {Meluzzi}},
  \ and\ \bibinfo {author} {\bibfnamefont {V.~A.}\ \bibnamefont
  {Mandelshtam}},\ }\href@noop {} {\bibfield  {journal} {\bibinfo  {journal}
  {Phys. Rev. Lett.}\ }\textbf {\bibinfo {volume} {96}},\ \bibinfo {eid}
  {113401} (\bibinfo {year} {2006})}\BibitemShut {NoStop}%
\bibitem [{\citenamefont {P{\'{e}}rez~de Tudela}\ \emph
  {et~al.}(2010)\citenamefont {P{\'{e}}rez~de Tudela}, \citenamefont
  {M{\'{a}}rquez-Mijares}, \citenamefont {Gonz{\'{a}}lez-Lezana}, \citenamefont
  {Roncero}, \citenamefont {Miret-Art{\'{e}}s}, \citenamefont
  {Delgado-Barrio},\ and\ \citenamefont {Villarreal}}]{Perez10a}%
  \BibitemOpen
  \bibfield  {author} {\bibinfo {author} {\bibfnamefont {R.}~\bibnamefont
  {P{\'{e}}rez~de Tudela}}, \bibinfo {author} {\bibfnamefont {M.}~\bibnamefont
  {M{\'{a}}rquez-Mijares}}, \bibinfo {author} {\bibfnamefont {T.}~\bibnamefont
  {Gonz{\'{a}}lez-Lezana}}, \bibinfo {author} {\bibfnamefont {O.}~\bibnamefont
  {Roncero}}, \bibinfo {author} {\bibfnamefont {S.}~\bibnamefont
  {Miret-Art{\'{e}}s}}, \bibinfo {author} {\bibfnamefont {G.}~\bibnamefont
  {Delgado-Barrio}}, \ and\ \bibinfo {author} {\bibfnamefont {P.}~\bibnamefont
  {Villarreal}},\ }\href@noop {} {\bibfield  {journal} {\bibinfo  {journal} {J.
  Chem. Phys.}\ }\textbf {\bibinfo {volume} {132}},\ \bibinfo {pages} {244303}
  (\bibinfo {year} {2010})}\BibitemShut {NoStop}%
\bibitem [{\citenamefont {Adjanor}\ \emph {et~al.}(2006)\citenamefont
  {Adjanor}, \citenamefont {Ath{\`e}nes},\ and\ \citenamefont
  {Calvo}}]{Adjanor2006a}%
  \BibitemOpen
  \bibfield  {author} {\bibinfo {author} {\bibfnamefont {G.}~\bibnamefont
  {Adjanor}}, \bibinfo {author} {\bibfnamefont {M.}~\bibnamefont
  {Ath{\`e}nes}}, \ and\ \bibinfo {author} {\bibfnamefont {F.}~\bibnamefont
  {Calvo}},\ }\href {\doibase 10.1140/epjb/e2006-00353-0} {\bibfield  {journal}
  {\bibinfo  {journal} {Eur. Phys. J. B}\ }\textbf {\bibinfo {volume} {53}},\
  \bibinfo {pages} {47} (\bibinfo {year} {2006})}\BibitemShut {NoStop}%
\bibitem [{\citenamefont {Pahl}\ \emph {et~al.}(2008)\citenamefont {Pahl},
  \citenamefont {Calvo}, \citenamefont {Ko\v{c}i},\ and\ \citenamefont
  {Schwerdtfeger}}]{Pahl2008a}%
  \BibitemOpen
  \bibfield  {author} {\bibinfo {author} {\bibfnamefont {E.}~\bibnamefont
  {Pahl}}, \bibinfo {author} {\bibfnamefont {F.}~\bibnamefont {Calvo}},
  \bibinfo {author} {\bibfnamefont {L.}~\bibnamefont {Ko\v{c}i}}, \ and\
  \bibinfo {author} {\bibfnamefont {P.}~\bibnamefont {Schwerdtfeger}},\
  }\href@noop {} {\bibfield  {journal} {\bibinfo  {journal} {Angew. Chem. Int.
  Ed.}\ }\textbf {\bibinfo {volume} {47}},\ \bibinfo {pages} {8207} (\bibinfo
  {year} {2008})}\BibitemShut {NoStop}%
\bibitem [{\citenamefont {Georgescu}\ and\ \citenamefont
  {Mandelshtam}(2011)}]{Georgescu2011a}%
  \BibitemOpen
  \bibfield  {author} {\bibinfo {author} {\bibfnamefont {I.}~\bibnamefont
  {Georgescu}}\ and\ \bibinfo {author} {\bibfnamefont {V.~A.}\ \bibnamefont
  {Mandelshtam}},\ }\href@noop {} {\bibfield  {journal} {\bibinfo  {journal}
  {J. Chem. Phys.}\ }\textbf {\bibinfo {volume} {135}},\ \bibinfo {eid}
  {154106} (\bibinfo {year} {2011})}\BibitemShut {NoStop}%
\bibitem [{\citenamefont {Georgescu}\ and\ \citenamefont
  {Mandelshtam}(2012)}]{Georgescu2012a}%
  \BibitemOpen
  \bibfield  {author} {\bibinfo {author} {\bibfnamefont {I.}~\bibnamefont
  {Georgescu}}\ and\ \bibinfo {author} {\bibfnamefont {V.~A.}\ \bibnamefont
  {Mandelshtam}},\ }\href@noop {} {\bibfield  {journal} {\bibinfo  {journal}
  {J. Chem. Phys.}\ }\textbf {\bibinfo {volume} {137}},\ \bibinfo {eid}
  {144106} (\bibinfo {year} {2012})}\BibitemShut {NoStop}%
\bibitem [{\citenamefont {Berne}\ and\ \citenamefont
  {Thirumalai}(1986)}]{Berne86a}%
  \BibitemOpen
  \bibfield  {author} {\bibinfo {author} {\bibfnamefont {B.~J.}\ \bibnamefont
  {Berne}}\ and\ \bibinfo {author} {\bibfnamefont {D.}~\bibnamefont
  {Thirumalai}},\ }\href@noop {} {\bibfield  {journal} {\bibinfo  {journal}
  {Annu. Rev. Phys. Chem.}\ }\textbf {\bibinfo {volume} {37}},\ \bibinfo
  {pages} {401} (\bibinfo {year} {1986})}\BibitemShut {NoStop}%
\bibitem [{\citenamefont {Makri}(1999)}]{Makri99a}%
  \BibitemOpen
  \bibfield  {author} {\bibinfo {author} {\bibfnamefont {N.}~\bibnamefont
  {Makri}},\ }\href@noop {} {\bibfield  {journal} {\bibinfo  {journal} {Annu.
  Rev. Phys. Chem.}\ }\textbf {\bibinfo {volume} {50}},\ \bibinfo {pages} {167}
  (\bibinfo {year} {1999})}\BibitemShut {NoStop}%
\bibitem [{\citenamefont {Ceperley}(2003)}]{Ceperley03a}%
  \BibitemOpen
  \bibfield  {author} {\bibinfo {author} {\bibfnamefont {D.~M.}\ \bibnamefont
  {Ceperley}},\ }\href@noop {} {\bibfield  {journal} {\bibinfo  {journal} {AIP
  Conf. Proc.}\ }\textbf {\bibinfo {volume} {690}},\ \bibinfo {pages} {85}
  (\bibinfo {year} {2003})}\BibitemShut {NoStop}%
\bibitem [{\citenamefont {Frantz}\ \emph {et~al.}(1992)\citenamefont {Frantz},
  \citenamefont {Freeman},\ and\ \citenamefont {Doll}}]{Frantz92a}%
  \BibitemOpen
  \bibfield  {author} {\bibinfo {author} {\bibfnamefont {D.~D.}\ \bibnamefont
  {Frantz}}, \bibinfo {author} {\bibfnamefont {D.~L.}\ \bibnamefont {Freeman}},
  \ and\ \bibinfo {author} {\bibfnamefont {J.~D.}\ \bibnamefont {Doll}},\
  }\href@noop {} {\bibfield  {journal} {\bibinfo  {journal} {J. Chem. Phys.}\
  }\textbf {\bibinfo {volume} {97}},\ \bibinfo {pages} {5713} (\bibinfo {year}
  {1992})}\BibitemShut {NoStop}%
\bibitem [{\citenamefont {Cartarius}\ and\ \citenamefont
  {Pollak}(2011)}]{Cartarius11a}%
  \BibitemOpen
  \bibfield  {author} {\bibinfo {author} {\bibfnamefont {H.}~\bibnamefont
  {Cartarius}}\ and\ \bibinfo {author} {\bibfnamefont {E.}~\bibnamefont
  {Pollak}},\ }\href@noop {} {\bibfield  {journal} {\bibinfo  {journal} {J.
  Chem. Phys.}\ }\textbf {\bibinfo {volume} {134}},\ \bibinfo {eid} {044107}
  (\bibinfo {year} {2011})}\BibitemShut {NoStop}%
\bibitem [{\citenamefont {Liu}\ and\ \citenamefont {Miller}(2011)}]{Liu2011a}%
  \BibitemOpen
  \bibfield  {author} {\bibinfo {author} {\bibfnamefont {J.}~\bibnamefont
  {Liu}}\ and\ \bibinfo {author} {\bibfnamefont {W.~H.}\ \bibnamefont
  {Miller}},\ }\href@noop {} {\bibfield  {journal} {\bibinfo  {journal} {J.
  Chem. Phys.}\ }\textbf {\bibinfo {volume} {134}},\ \bibinfo {eid} {104102}
  (\bibinfo {year} {2011})}\BibitemShut {NoStop}%
\bibitem [{\citenamefont {Cartarius}\ and\ \citenamefont
  {Pollak}(2012)}]{Cartarius12a}%
  \BibitemOpen
  \bibfield  {author} {\bibinfo {author} {\bibfnamefont {H.}~\bibnamefont
  {Cartarius}}\ and\ \bibinfo {author} {\bibfnamefont {E.}~\bibnamefont
  {Pollak}},\ }\href@noop {} {\bibfield  {journal} {\bibinfo  {journal} {Chem.
  Phys.}\ }\textbf {\bibinfo {volume} {399}},\ \bibinfo {pages} {135} (\bibinfo
  {year} {2012})}\BibitemShut {NoStop}%
\bibitem [{\citenamefont {Kryvohuz}(2012)}]{Kryvohuz2012a}%
  \BibitemOpen
  \bibfield  {author} {\bibinfo {author} {\bibfnamefont {M.}~\bibnamefont
  {Kryvohuz}},\ }\href@noop {} {\bibfield  {journal} {\bibinfo  {journal}
  {Chem. Phys.}\ }\textbf {\bibinfo {volume} {407}},\ \bibinfo {pages} {124 }
  (\bibinfo {year} {2012})}\BibitemShut {NoStop}%
\bibitem [{\citenamefont {Conte}\ and\ \citenamefont
  {Pollak}(2012)}]{Conte2012a}%
  \BibitemOpen
  \bibfield  {author} {\bibinfo {author} {\bibfnamefont {R.}~\bibnamefont
  {Conte}}\ and\ \bibinfo {author} {\bibfnamefont {E.}~\bibnamefont {Pollak}},\
  }\href@noop {} {\bibfield  {journal} {\bibinfo  {journal} {J. Chem. Phys.}\
  }\textbf {\bibinfo {volume} {136}},\ \bibinfo {eid} {094101} (\bibinfo {year}
  {2012})}\BibitemShut {NoStop}%
\bibitem [{\citenamefont {Etters}\ and\ \citenamefont
  {Kaelberer}(1975)}]{Etters75a}%
  \BibitemOpen
  \bibfield  {author} {\bibinfo {author} {\bibfnamefont {R.~D.}\ \bibnamefont
  {Etters}}\ and\ \bibinfo {author} {\bibfnamefont {J.}~\bibnamefont
  {Kaelberer}},\ }\href@noop {} {\bibfield  {journal} {\bibinfo  {journal}
  {Phys. Rev. A}\ }\textbf {\bibinfo {volume} {11}},\ \bibinfo {pages} {1068}
  (\bibinfo {year} {1975})}\BibitemShut {NoStop}%
\bibitem [{\citenamefont {Leitner}\ \emph {et~al.}(1989)\citenamefont
  {Leitner}, \citenamefont {Berry},\ and\ \citenamefont
  {Whitnell}}]{Leitner89a}%
  \BibitemOpen
  \bibfield  {author} {\bibinfo {author} {\bibfnamefont {D.~M.}\ \bibnamefont
  {Leitner}}, \bibinfo {author} {\bibfnamefont {R.~S.}\ \bibnamefont {Berry}},
  \ and\ \bibinfo {author} {\bibfnamefont {R.~M.}\ \bibnamefont {Whitnell}},\
  }\href@noop {} {\bibfield  {journal} {\bibinfo  {journal} {J. Chem. Phys.}\
  }\textbf {\bibinfo {volume} {91}},\ \bibinfo {pages} {3470} (\bibinfo {year}
  {1989})}\BibitemShut {NoStop}%
\bibitem [{\citenamefont {Leitner}\ \emph {et~al.}(1991)\citenamefont
  {Leitner}, \citenamefont {Doll},\ and\ \citenamefont
  {Whitnell}}]{Leitner91a}%
  \BibitemOpen
  \bibfield  {author} {\bibinfo {author} {\bibfnamefont {D.~M.}\ \bibnamefont
  {Leitner}}, \bibinfo {author} {\bibfnamefont {J.~D.}\ \bibnamefont {Doll}}, \
  and\ \bibinfo {author} {\bibfnamefont {R.~M.}\ \bibnamefont {Whitnell}},\
  }\href@noop {} {\bibfield  {journal} {\bibinfo  {journal} {J. Chem. Phys.}\
  }\textbf {\bibinfo {volume} {94}},\ \bibinfo {pages} {6644} (\bibinfo {year}
  {1991})}\BibitemShut {NoStop}%
\bibitem [{\citenamefont {Elyutin}\ \emph {et~al.}(1994)\citenamefont
  {Elyutin}, \citenamefont {Baranov}, \citenamefont {Belega},\ and\
  \citenamefont {Trubnikov}}]{Elyutin94a}%
  \BibitemOpen
  \bibfield  {author} {\bibinfo {author} {\bibfnamefont {P.~V.}\ \bibnamefont
  {Elyutin}}, \bibinfo {author} {\bibfnamefont {V.~I.}\ \bibnamefont
  {Baranov}}, \bibinfo {author} {\bibfnamefont {E.~D.}\ \bibnamefont {Belega}},
  \ and\ \bibinfo {author} {\bibfnamefont {D.~N.}\ \bibnamefont {Trubnikov}},\
  }\href@noop {} {\bibfield  {journal} {\bibinfo  {journal} {J. Chem. Phys.}\
  }\textbf {\bibinfo {volume} {100}},\ \bibinfo {pages} {3843} (\bibinfo {year}
  {1994})}\BibitemShut {NoStop}%
\bibitem [{\citenamefont {Gonz{\'{a}}lez-Lezana}\ \emph
  {et~al.}(1999)\citenamefont {Gonz{\'{a}}lez-Lezana}, \citenamefont
  {Rubayo-Soneira}, \citenamefont {Miret-Art{\'{e}}s}, \citenamefont
  {Gianturco}, \citenamefont {Delgado-Barrio},\ and\ \citenamefont
  {Villarreal}}]{Gonzales-Lezana99a}%
  \BibitemOpen
  \bibfield  {author} {\bibinfo {author} {\bibfnamefont {T.}~\bibnamefont
  {Gonz{\'{a}}lez-Lezana}}, \bibinfo {author} {\bibfnamefont {J.}~\bibnamefont
  {Rubayo-Soneira}}, \bibinfo {author} {\bibfnamefont {S.}~\bibnamefont
  {Miret-Art{\'{e}}s}}, \bibinfo {author} {\bibfnamefont {F.~A.}\ \bibnamefont
  {Gianturco}}, \bibinfo {author} {\bibfnamefont {G.}~\bibnamefont
  {Delgado-Barrio}}, \ and\ \bibinfo {author} {\bibfnamefont {P.}~\bibnamefont
  {Villarreal}},\ }\href@noop {} {\bibfield  {journal} {\bibinfo  {journal} {J.
  Chem. Phys.}\ }\textbf {\bibinfo {volume} {110}},\ \bibinfo {pages} {9000}
  (\bibinfo {year} {1999})}\BibitemShut {NoStop}%
\bibitem [{\citenamefont {Svr{\v c}kov{\'a}}\ \emph {et~al.}(2011)\citenamefont
  {Svr{\v c}kov{\'a}}, \citenamefont {V{\'i}tek}, \citenamefont {Karlick{\'y}},
  \citenamefont {Paidarov{\'a}},\ and\ \citenamefont {Kalus}}]{Svrckova2011a}%
  \BibitemOpen
  \bibfield  {author} {\bibinfo {author} {\bibfnamefont {P.}~\bibnamefont
  {Svr{\v c}kov{\'a}}}, \bibinfo {author} {\bibfnamefont {A.}~\bibnamefont
  {V{\'i}tek}}, \bibinfo {author} {\bibfnamefont {F.}~\bibnamefont
  {Karlick{\'y}}}, \bibinfo {author} {\bibfnamefont {I.}~\bibnamefont
  {Paidarov{\'a}}}, \ and\ \bibinfo {author} {\bibfnamefont {R.}~\bibnamefont
  {Kalus}},\ }\href {\doibase http://dx.doi.org/10.1063/1.3599052} {\bibfield
  {journal} {\bibinfo  {journal} {J. Chem. Phys.}\ }\textbf {\bibinfo {volume}
  {134}},\ \bibinfo {eid} {224310} (\bibinfo {year} {2011})}\BibitemShut
  {NoStop}%
\bibitem [{\citenamefont {Blanco}\ and\ \citenamefont
  {García}(2013)}]{Blanco2013a}%
  \BibitemOpen
  \bibfield  {author} {\bibinfo {author} {\bibfnamefont {F.}~\bibnamefont
  {Blanco}}\ and\ \bibinfo {author} {\bibfnamefont {G.}~\bibnamefont
  {García}},\ }\href {http://stacks.iop.org/1742-6596/438/i=1/a=012012}
  {\bibfield  {journal} {\bibinfo  {journal} {J. Phys.: Conf. Ser.}\ }\textbf
  {\bibinfo {volume} {438}},\ \bibinfo {pages} {012012} (\bibinfo {year}
  {2013})}\BibitemShut {NoStop}%
\bibitem [{\citenamefont {Unn-Toc}\ \emph {et~al.}(2012)\citenamefont
  {Unn-Toc}, \citenamefont {Halberstadt}, \citenamefont {Meier},\ and\
  \citenamefont {Mella}}]{UnnToc2012a}%
  \BibitemOpen
  \bibfield  {author} {\bibinfo {author} {\bibfnamefont {W.}~\bibnamefont
  {Unn-Toc}}, \bibinfo {author} {\bibfnamefont {N.}~\bibnamefont
  {Halberstadt}}, \bibinfo {author} {\bibfnamefont {C.}~\bibnamefont {Meier}},
  \ and\ \bibinfo {author} {\bibfnamefont {M.}~\bibnamefont {Mella}},\ }\href
  {\doibase http://dx.doi.org/10.1063/1.4730033} {\bibfield  {journal}
  {\bibinfo  {journal} {J. Chem. Phys.}\ }\textbf {\bibinfo {volume} {137}},\
  \bibinfo {eid} {014304} (\bibinfo {year} {2012})}\BibitemShut {NoStop}%
\bibitem [{\citenamefont {Mella}(2009)}]{Mella2009a}%
  \BibitemOpen
  \bibfield  {author} {\bibinfo {author} {\bibfnamefont {M.}~\bibnamefont
  {Mella}},\ }\href {\doibase http://dx.doi.org/10.1063/1.3239476} {\bibfield
  {journal} {\bibinfo  {journal} {J. Chem. Phys.}\ }\textbf {\bibinfo {volume}
  {131}},\ \bibinfo {eid} {124309} (\bibinfo {year} {2009})}\BibitemShut
  {NoStop}%
\bibitem [{\citenamefont {Calvo}\ \emph {et~al.}(2012)\citenamefont {Calvo},
  \citenamefont {Naumkin},\ and\ \citenamefont {Wales}}]{Calvo2012a}%
  \BibitemOpen
  \bibfield  {author} {\bibinfo {author} {\bibfnamefont {F.}~\bibnamefont
  {Calvo}}, \bibinfo {author} {\bibfnamefont {F.}~\bibnamefont {Naumkin}}, \
  and\ \bibinfo {author} {\bibfnamefont {D.}~\bibnamefont {Wales}},\ }\href
  {\doibase http://dx.doi.org/10.1016/j.cplett.2012.09.013} {\bibfield
  {journal} {\bibinfo  {journal} {Chem. Phys. Lett.}\ }\textbf {\bibinfo
  {volume} {551}},\ \bibinfo {pages} {38 } (\bibinfo {year}
  {2012})}\BibitemShut {NoStop}%
\bibitem [{\citenamefont {M{\"a}hr}\ \emph {et~al.}(2007)\citenamefont
  {M{\"a}hr}, \citenamefont {Zappa}, \citenamefont {Denifl}, \citenamefont
  {Kubala}, \citenamefont {Echt}, \citenamefont {M{\"a}rk},\ and\ \citenamefont
  {Scheier}}]{Maehr2007a}%
  \BibitemOpen
  \bibfield  {author} {\bibinfo {author} {\bibfnamefont {I.}~\bibnamefont
  {M{\"a}hr}}, \bibinfo {author} {\bibfnamefont {F.}~\bibnamefont {Zappa}},
  \bibinfo {author} {\bibfnamefont {S.}~\bibnamefont {Denifl}}, \bibinfo
  {author} {\bibfnamefont {D.}~\bibnamefont {Kubala}}, \bibinfo {author}
  {\bibfnamefont {O.}~\bibnamefont {Echt}}, \bibinfo {author} {\bibfnamefont
  {T.~D.}\ \bibnamefont {M{\"a}rk}}, \ and\ \bibinfo {author} {\bibfnamefont
  {P.}~\bibnamefont {Scheier}},\ }\href {\doibase
  10.1103/PhysRevLett.98.023401} {\bibfield  {journal} {\bibinfo  {journal}
  {Phys. Rev. Lett.}\ }\textbf {\bibinfo {volume} {98}},\ \bibinfo {pages}
  {023401} (\bibinfo {year} {2007})}\BibitemShut {NoStop}%
\bibitem [{\citenamefont {Calvo}\ and\ \citenamefont
  {Parneix}(2009)}]{Calvo2009a}%
  \BibitemOpen
  \bibfield  {author} {\bibinfo {author} {\bibfnamefont {F.}~\bibnamefont
  {Calvo}}\ and\ \bibinfo {author} {\bibfnamefont {P.}~\bibnamefont
  {Parneix}},\ }\href {\doibase 10.1021/jp903282b} {\bibfield  {journal}
  {\bibinfo  {journal} {J. Phys. Chem. A}\ }\textbf {\bibinfo {volume} {113}},\
  \bibinfo {pages} {14352} (\bibinfo {year} {2009})}\BibitemShut {NoStop}%
\bibitem [{\citenamefont {Franke}\ \emph {et~al.}(1993)\citenamefont {Franke},
  \citenamefont {Hilf},\ and\ \citenamefont {Borrmann}}]{Franke1993a}%
  \BibitemOpen
  \bibfield  {author} {\bibinfo {author} {\bibfnamefont {G.}~\bibnamefont
  {Franke}}, \bibinfo {author} {\bibfnamefont {E.~R.}\ \bibnamefont {Hilf}}, \
  and\ \bibinfo {author} {\bibfnamefont {P.}~\bibnamefont {Borrmann}},\ }\href
  {\doibase http://dx.doi.org/10.1063/1.464070} {\bibfield  {journal} {\bibinfo
   {journal} {J. Chem. Phys.}\ }\textbf {\bibinfo {volume} {98}},\ \bibinfo
  {pages} {3496} (\bibinfo {year} {1993})}\BibitemShut {NoStop}%
\bibitem [{\citenamefont {Shao}\ and\ \citenamefont {Pollak}(2006)}]{Shao06a}%
  \BibitemOpen
  \bibfield  {author} {\bibinfo {author} {\bibfnamefont {J.}~\bibnamefont
  {Shao}}\ and\ \bibinfo {author} {\bibfnamefont {E.}~\bibnamefont {Pollak}},\
  }\href@noop {} {\bibfield  {journal} {\bibinfo  {journal} {J. Chem. Phys.}\
  }\textbf {\bibinfo {volume} {125}},\ \bibinfo {eid} {133502} (\bibinfo {year}
  {2006})}\BibitemShut {NoStop}%
\bibitem [{\citenamefont {Zhang}\ \emph {et~al.}(2009)\citenamefont {Zhang},
  \citenamefont {Shao},\ and\ \citenamefont {Pollak}}]{Zhang09a}%
  \BibitemOpen
  \bibfield  {author} {\bibinfo {author} {\bibfnamefont {D.~H.}\ \bibnamefont
  {Zhang}}, \bibinfo {author} {\bibfnamefont {J.}~\bibnamefont {Shao}}, \ and\
  \bibinfo {author} {\bibfnamefont {E.}~\bibnamefont {Pollak}},\ }\href@noop {}
  {\bibfield  {journal} {\bibinfo  {journal} {J. Chem. Phys.}\ }\textbf
  {\bibinfo {volume} {131}},\ \bibinfo {pages} {044116} (\bibinfo {year}
  {2009})}\BibitemShut {NoStop}%
\bibitem [{\citenamefont {Kwon}\ and\ \citenamefont
  {Moscowitz}(1996)}]{Kwon1996a}%
  \BibitemOpen
  \bibfield  {author} {\bibinfo {author} {\bibfnamefont {K.}~\bibnamefont
  {Kwon}}\ and\ \bibinfo {author} {\bibfnamefont {A.}~\bibnamefont
  {Moscowitz}},\ }\href {\doibase 10.1103/PhysRevLett.77.1238} {\bibfield
  {journal} {\bibinfo  {journal} {Phys. Rev. Lett.}\ }\textbf {\bibinfo
  {volume} {77}},\ \bibinfo {pages} {1238} (\bibinfo {year}
  {1996})}\BibitemShut {NoStop}%
\bibitem [{\citenamefont {Ulrich}\ \emph {et~al.}(2011)\citenamefont {Ulrich},
  \citenamefont {Vredenborg}, \citenamefont {Malakzadeh}, \citenamefont
  {Schmidt}, \citenamefont {Havermeier}, \citenamefont {Meckel}, \citenamefont
  {Cole}, \citenamefont {Smolarski}, \citenamefont {Chang}, \citenamefont
  {Jahnke},\ and\ \citenamefont {D{\"o}rner}}]{Ulrich2011a}%
  \BibitemOpen
  \bibfield  {author} {\bibinfo {author} {\bibfnamefont {B.}~\bibnamefont
  {Ulrich}}, \bibinfo {author} {\bibfnamefont {A.}~\bibnamefont {Vredenborg}},
  \bibinfo {author} {\bibfnamefont {A.}~\bibnamefont {Malakzadeh}}, \bibinfo
  {author} {\bibfnamefont {L.~P.~H.}\ \bibnamefont {Schmidt}}, \bibinfo
  {author} {\bibfnamefont {T.}~\bibnamefont {Havermeier}}, \bibinfo {author}
  {\bibfnamefont {M.}~\bibnamefont {Meckel}}, \bibinfo {author} {\bibfnamefont
  {K.}~\bibnamefont {Cole}}, \bibinfo {author} {\bibfnamefont {M.}~\bibnamefont
  {Smolarski}}, \bibinfo {author} {\bibfnamefont {Z.}~\bibnamefont {Chang}},
  \bibinfo {author} {\bibfnamefont {T.}~\bibnamefont {Jahnke}}, \ and\ \bibinfo
  {author} {\bibfnamefont {R.}~\bibnamefont {D{\"o}rner}},\ }\href@noop {}
  {\bibfield  {journal} {\bibinfo  {journal} {J. Phys. Chem. A}\ }\textbf
  {\bibinfo {volume} {115}},\ \bibinfo {pages} {6936} (\bibinfo {year}
  {2011})}\BibitemShut {NoStop}%
\bibitem [{\citenamefont {Conte}\ and\ \citenamefont
  {Pollak}(2010)}]{Conte10a}%
  \BibitemOpen
  \bibfield  {author} {\bibinfo {author} {\bibfnamefont {R.}~\bibnamefont
  {Conte}}\ and\ \bibinfo {author} {\bibfnamefont {E.}~\bibnamefont {Pollak}},\
  }\href@noop {} {\bibfield  {journal} {\bibinfo  {journal} {Phys. Rev. E}\
  }\textbf {\bibinfo {volume} {81}},\ \bibinfo {pages} {036704} (\bibinfo
  {year} {2010})}\BibitemShut {NoStop}%
\bibitem [{\citenamefont {Frantsuzov}\ \emph {et~al.}(2003)\citenamefont
  {Frantsuzov}, \citenamefont {Neumaier},\ and\ \citenamefont
  {Mandelshtam}}]{Frantsuzov03a}%
  \BibitemOpen
  \bibfield  {author} {\bibinfo {author} {\bibfnamefont {P.}~\bibnamefont
  {Frantsuzov}}, \bibinfo {author} {\bibfnamefont {A.}~\bibnamefont
  {Neumaier}}, \ and\ \bibinfo {author} {\bibfnamefont {V.~A.}\ \bibnamefont
  {Mandelshtam}},\ }\href@noop {} {\bibfield  {journal} {\bibinfo  {journal}
  {Chem. Phys. Lett.}\ }\textbf {\bibinfo {volume} {381}},\ \bibinfo {pages}
  {117} (\bibinfo {year} {2003})}\BibitemShut {NoStop}%
\bibitem [{\citenamefont {Aziz}\ and\ \citenamefont {Slaman}(1985)}]{Aziz86a}%
  \BibitemOpen
  \bibfield  {author} {\bibinfo {author} {\bibfnamefont {R.~A.}\ \bibnamefont
  {Aziz}}\ and\ \bibinfo {author} {\bibfnamefont {M.~J.}\ \bibnamefont
  {Slaman}},\ }\href@noop {} {\bibfield  {journal} {\bibinfo  {journal} {Mol.
  Phys.}\ }\textbf {\bibinfo {volume} {58}},\ \bibinfo {pages} {679} (\bibinfo
  {year} {1985})}\BibitemShut {NoStop}%
\bibitem [{\citenamefont {Frantsuzov}\ and\ \citenamefont
  {Mandelshtam}(2008)}]{Frantsuzov08a}%
  \BibitemOpen
  \bibfield  {author} {\bibinfo {author} {\bibfnamefont {P.~A.}\ \bibnamefont
  {Frantsuzov}}\ and\ \bibinfo {author} {\bibfnamefont {V.~A.}\ \bibnamefont
  {Mandelshtam}},\ }\href@noop {} {\bibfield  {journal} {\bibinfo  {journal}
  {J. Chem. Phys.}\ }\textbf {\bibinfo {volume} {128}},\ \bibinfo {pages}
  {094304} (\bibinfo {year} {2008})}\BibitemShut {NoStop}%
\bibitem [{\citenamefont {Franke}\ \emph {et~al.}(1988)\citenamefont {Franke},
  \citenamefont {Hilf},\ and\ \citenamefont {Polley}}]{Franke88a}%
  \BibitemOpen
  \bibfield  {author} {\bibinfo {author} {\bibfnamefont {G.}~\bibnamefont
  {Franke}}, \bibinfo {author} {\bibfnamefont {E.}~\bibnamefont {Hilf}}, \ and\
  \bibinfo {author} {\bibfnamefont {L.}~\bibnamefont {Polley}},\ }\href@noop {}
  {\bibfield  {journal} {\bibinfo  {journal} {Z. Phys. D}\ }\textbf {\bibinfo
  {volume} {9}},\ \bibinfo {pages} {343} (\bibinfo {year} {1988})}\BibitemShut
  {NoStop}%
\bibitem [{\citenamefont {Borrmann}(1994)}]{Borrmann94a}%
  \BibitemOpen
  \bibfield  {author} {\bibinfo {author} {\bibfnamefont {P.}~\bibnamefont
  {Borrmann}},\ }\href@noop {} {\bibfield  {journal} {\bibinfo  {journal}
  {Comput. Mater. Sci.}\ }\textbf {\bibinfo {volume} {2}},\ \bibinfo {pages}
  {593} (\bibinfo {year} {1994})}\BibitemShut {NoStop}%
\bibitem [{\citenamefont {Tsai}\ and\ \citenamefont {Jordan}(1993)}]{Tsai93a}%
  \BibitemOpen
  \bibfield  {author} {\bibinfo {author} {\bibfnamefont {C.~J.}\ \bibnamefont
  {Tsai}}\ and\ \bibinfo {author} {\bibfnamefont {K.~D.}\ \bibnamefont
  {Jordan}},\ }\href@noop {} {\bibfield  {journal} {\bibinfo  {journal} {J.
  Chem. Phys.}\ }\textbf {\bibinfo {volume} {99}},\ \bibinfo {pages} {6957}
  (\bibinfo {year} {1993})}\BibitemShut {NoStop}%
\end{thebibliography}
\end{document}